\documentclass[aps,prd,preprintnumbers,showpacs,showkeys,nofootinbib,
superscriptaddress,fleqn,floatfix,tightenlines,10pt]{revtex4-1}
\usepackage{amsmath,amsfonts,amssymb,amscd,amsxtra,amsthm}
\usepackage{graphicx}   
\usepackage{epstopdf}
\usepackage{dcolumn}    
\usepackage{bm}         
\usepackage{slashed}
\usepackage{cancel}
\usepackage{float}
\usepackage{mathtools}
\usepackage{amsbsy}
\usepackage{amstext}
\usepackage[toc,page]{appendix}

\usepackage[utf8]{inputenc}
\usepackage{booktabs}
\usepackage[normalem]{ulem} 
\usepackage[dvipsnames]{xcolor} 
\usepackage{tabularx}
\usepackage{enumitem}
\usepackage{array}
\usepackage{slashed}
\usepackage{tikz}
\usepackage{float}
\usepackage{multirow}
\renewcommand\sout{\bgroup \color{red} \ULdepth=-.5ex \ULset}

\begin{document}
\preprint{PKNU-NuHaTh-2020-01}
\title{Investigation of electroproduction of $\phi$ mesons off protons}
\author{Sang-Ho Kim}
\email[E-mail: ]{shkim@pknu.ac.kr}
\affiliation{Department of Physics, Pukyong National University (PKNU),
Busan 48513, Republic of Korea}
\author{Seung-il Nam}
\email[E-mail: ]{sinam@pknu.ac.kr}
\affiliation{Department of Physics, Pukyong National University (PKNU),
Busan 48513, Republic of Korea}
\affiliation{Asia Pacific Center for Theoretical Physics (APCTP), Pohang
37673, Republic of Korea}
\date{\today}
\begin{abstract}
We investigate $\phi$-meson electroproduction off the proton target, i.e.,
$\gamma^* p \to \phi p$, by employing a tree-level effective Lagrangian approach
in the kinematical ranges of $Q^2$ = (0$-$4) $\mathrm{GeV}^2$, $W$ = (2$-$5) GeV,
and $|t| \leq 2\,\mathrm{GeV}^2$.
In addition to the universally accepted Pomeron exchange, we consider various
meson exchanges in the $t$ channel with the Regge method.
Direct $\phi$-meson radiations in the $s$- and $u$-channels are also taken into
account.
We find that the $Q^2$ dependence of the transverse ($\sigma_{\mathrm{T}}$) and
longitudinal ($\sigma_{\mathrm{L}}$) cross sections are governed by Pomeron and
($a_0,f_0$) scalar meson exchanges, respectively.
Meanwhile, the contributions of $(\pi,\eta)$ pseudoscalar- and $f_1(1285)$
axial-vector-meson exchanges are much more suppressed.
The results of the interference cross sections ($\sigma_{\mathrm{LT}},
\sigma_{\mathrm{LT}}$) and the spin-density matrix elements indicate that
$s$-channel helicity conservation holds at $Q^2$ = (1$-$4)
$\mathrm{GeV}^2$.
The result of the parity asymmetry yield $P \simeq 0.95$ at $W$ = 2.5 GeV,
meaning that natural-parity exchange dominates the reaction process.
Our numerical results are in fair agreement with the experimental data and thus
the use of our effective Reggeized model is justified over the considered
kinematical ranges of $Q^2$, $W$, and $t$.
\end{abstract}
\pacs{13.40.-f, 13.60.Le, 14.40.Cs}
\keywords{$\phi$-meson electroproduction, Pomeron, $(a_0,f_0)$ scalar meson,
effective Lagrangian approach, Regge method.}
\maketitle

\section{Introduction}
Exclusive electroproduction of vector mesons is a suitable place to test model
predictions in a kinematic region where the transition between the hadronic and
partonic domains is involved according to the ranges of the photon virtuality
$Q^2$ and the photon center-of-mass (c.m.) energies $W$.
The ZEUS and H1 Collaborations at HERA accumulated a lot of data for
electroproductions of $\rho$-~\cite{Breitweg:1998nh,Aaron:2009xp}, $\omega$-~
\cite{Breitweg:2000mu}, and $\phi$-~\cite{Adloff:2000nx,Chekanov:2005cqa,
Aaron:2009xp} light vector-mesons over wide ranges of $Q^2$ and $W$
(e.g., $2.5 \leq Q^2 \leq 60$ $\mathrm{GeV}^2$ and $35 \leq W \leq 180$
GeV at H1~\cite{Aaron:2009xp}).
The scale is large enough for perturbative quantum chromodynamics (pQCD) to
be employed.
Meanwhile, the Cornell group at the Laboratory of Nuclear Studies (LNS) at
Cornell University ~\cite{Dixon:1977pg,Dixon:1978vy,Cassel:1981sx}, the CLAS
Collaboration at Jefferson Lab ~\cite{Hadjidakis:2004zm,Morrow:2008ek,
Lukashin:2001sh,Morand:2005ex,Santoro:2008ai}, and the HERMES Collaboration
at DESY~\cite{Borissov:2000wb,Airapetian:2010dh,Airapetian:2014gfp,
Airapetian:2017vit}
performed the experiments of vector-meson electroproductions at relatively low
$Q^2$ and $W$ values (e.g., $1.0 \leq Q^2 \leq 7.0$ $\mathrm{GeV}^2$ and $3.0
\leq W \leq 6.3$ GeV at HERMES).
Then we get access to the soft scale and pose a question.
What ranges in $Q^2$ and $W$ are more adequate for the hadronic or partonic
descriptions?

A series of works on electro- and photoproductions of light vector-mesons was
carried out by Laget {\it et al.} previously based on the Regge phenomenology
~\cite{Laget:1994ba,Laget:2000gj,Laget:2001mu,Laget:2004qu}.
The exchanges of meson Regge trajectories and of a Pomeron trajectory are
considered in the $t$ channel with phenomenological form factors for each
vertex.
The $Q^2$ and $t$ dependences on the cross sections at low photon energies
($W\approx \mathrm{a\,few}\,\,\mathrm{GeV}$) are reasonably described.
Then Ref.~~\cite{Obukhovsky:2009th} developed the work for $\rho$-meson
electroproduction by employing an effective Lagrangian approach with the
updated CLAS data~\cite{Hadjidakis:2004zm,Morrow:2008ek}.
The transverse and longitudinal parts of the cross sections are examined in
some detail.
The separated components help us to pin down the role of different meson
exchanges and Pomeron exchange as well, which is difficult only with the study
of the unpolarized total cross section.

In this paper, we take a similar approach for $\phi$-meson electroproduction
and test whether our hadronic description is applicable or not in the
kinematical ranges of $Q^2$ = (0$-$4) $\mathrm{GeV}^2$, $W$ = (2$-$5) GeV, and
$|t| \leq 2\,\mathrm{GeV}^2$.
We examine the $Q^2$ and $t$ dependences on the transverse, longitudinal, and
interference parts of the cross sections.
The latter enables us to test $s$-channel helicity conservation (SCHC).
The spin-density matrix elements of the produced $\phi$ meson are also analyzed
in the helicity frame which is in favor of SCHC. 
Parity asymmetry is calculated to check the relative strengths of natural to
unnatural parity exchanges in the $t$ channel.

For this purpose, we utilize our recent results for $\phi$-meson
photoproduction, $\gamma p \to \phi p$~\cite{Kim:2019kef}, where 
the relative contributions among the Pomeron and various meson exchanges were
discussed in detail by analyzing a vast amount of CLAS data
~\cite{Seraydaryan:2013ija,Dey:2014tfa}.
The basic formalism used in Ref.~\cite{Kim:2019kef} applied to the present work.
However, the $s$-channel nucleon resonance contribution is excluded for brevity,
although our kinematical range covers some high-mass resonance regions.
Indeed, in $\phi$-meson photoproduction, the $N^*$ contribution is found to
be crucial only for the backward $\phi$-meson scattering angles with small
magnitudes and does not change much the integrated cross sections
~\cite{Kim:2019kef}, whereas the data for electro- and photoproductions of
$\rho$-~\cite{Obukhovsky:2009th,Oh:2003aw} and $\omega$-~\cite{Morand:2005ex,
Wei:2019imo} vector-mesons imply the necessity for the $s$-channel $N^*$
contribution.
We find that our hadronic approach provides a very successful description of
the available experimental data over the considered kinematical ranges of $Q^2$,
$W$, and $t$.

The remaining part of this paper is organized as follows.
In Sec.~\ref{SecII}, we define the kinematics of $\phi$-meson electroproduction
process.
In Sec.~\ref{SecIII}, we explain the general formalism of the effective
Lagrangian approach.
We present and discuss the numerical results in Sec.~\ref{SecIV}.
The final section is devoted to the summary.

\section{Kinematics}
\label{SecII}
\begin{figure}[htp]
\includegraphics[width=8cm]{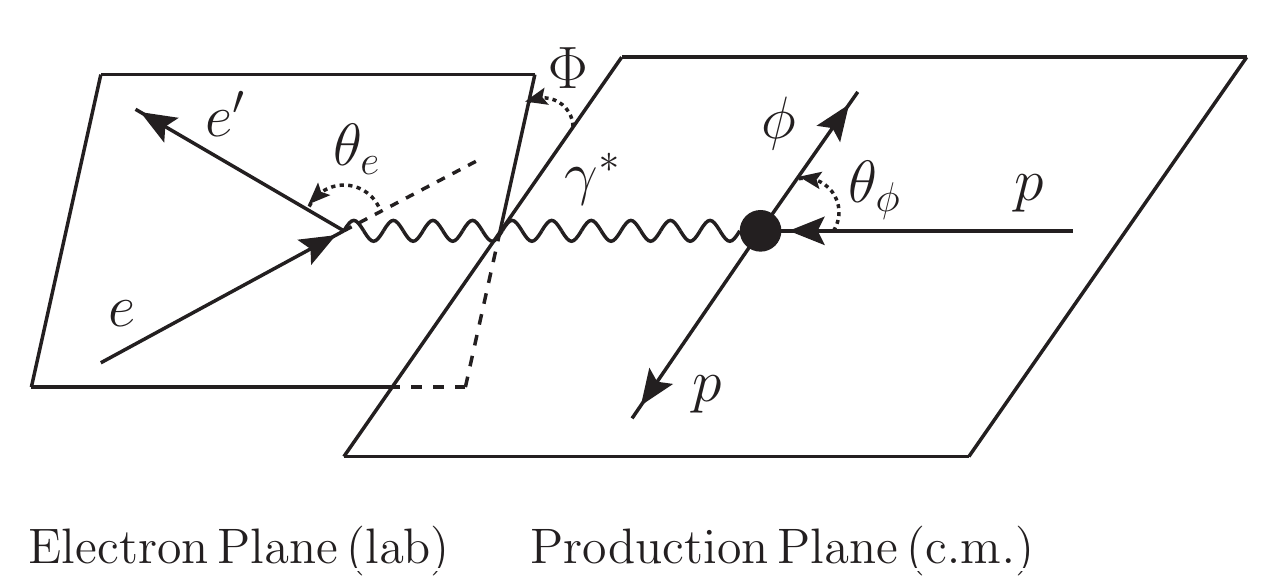}
\caption{Graphical representation of the electron scattering ($e e' \gamma^*$)-
and hadron production ($\gamma^* \phi p $)-planes for the $e p \to e'\phi p$
reaction defined in the laboratory (lab) and $\gamma^* p$ center-of-mass (c.m.)
frames, respectively.}
\label{FIG1:RP}
\end{figure}

Let us first specify kinematics of the $\phi$-electroproduction process
$e p \to e \phi p$ drawn in Fig.~\ref{FIG1:RP} graphically.
The four-momenta of the involved particles described in the hadron
production plane are given by
\begin{align}
\label{eq:PhiElecPro}
\gamma^*(k_1) + p(p_1) \to \phi(k_2) + p(p_2) ,
\end{align}
in parentheses, where
\begin{align}
\label{eq:def:FourMomenta}
& k_1 = (\sqrt{k^2-Q^2},\, 0,\, 0,\, k), \hspace{1em}
  k_2 = (\sqrt{p^2+M_\phi^2},\, p\sin\theta_\phi,\, 0,\, p\cos\theta_\phi),   \cr
& p_1 = (\sqrt{k^2+M_N^2},\, 0,\, 0,\, -k), \hspace{1em}
  p_2 = (\sqrt{p^2+M_N^2},\, -p\sin\theta_\phi,\, 0,\, -p\cos\theta_\phi) .
\end{align}
They are defined in the $\gamma^* p$ center-of-mass (c.m.) frame where the
$z$ axis is set to be parallel to the direction of the virtual photon and
the $y$ axis normal to the hadron production plane along
$\vec k_1 \times \vec k_2$.
Here, the magnitudes of the three-momenta of the initial- and
final-particles are given by
\begin{align}
\label{eq:def:kp}
k =& \lambda (-Q^2,M_N^2,W^2)^{\frac12} / (2W)
= M_N\sqrt{\nu^2+Q^2}/W,                                                  \cr
p =& \lambda (M_\phi^2,M_N^2,W^2)^{\frac12} / (2W),
\end{align}
where the K\"all\'en function is defined as
$\lambda (x,y,z) \equiv x^{2}+y^{2}+z^{2}-2(xy+yz+zx)$.

We define some relevant variables as follows:
\begin{itemize}
\item $Q^2 = -k_1^2 > 0$, the negative four-momentum squared of the virtual
photon, i.e., photon virtuality;
\item $W^2 = (k_1+p_1)^2 = M_N^2+2M_N\nu-Q^2$, the square of the invariant mass of
the $\gamma^* p$ system, where $\nu =  E_e - E_{e'}$ is the energy transfer from
the incident electron to the virtual photon in the laboratory (lab) frame;
\item $t=(k_1 - k_2)^2$, the squared four-momentum transfer from the $\gamma^*$
to the $\phi$.
$t' = |t-t_{min}|$, $t_{min}$ being the minimal value of $t$ at fixed $Q^2$ and
$W$;
\item $\theta_e$, the angle between the incident and scattered electrons;
\item $\Phi$, the angle between the electron scattering ($e e' \gamma^*$) and
hadron production ($\gamma^* \phi p$) planes;
\item $\theta_\phi$, the c.m. $\phi$-meson angle relative to the virtual photon
direction;
\end{itemize}

\section{Theoretical framework}
\label{SecIII}
\begin{figure}[ht]
\includegraphics[width=6cm]{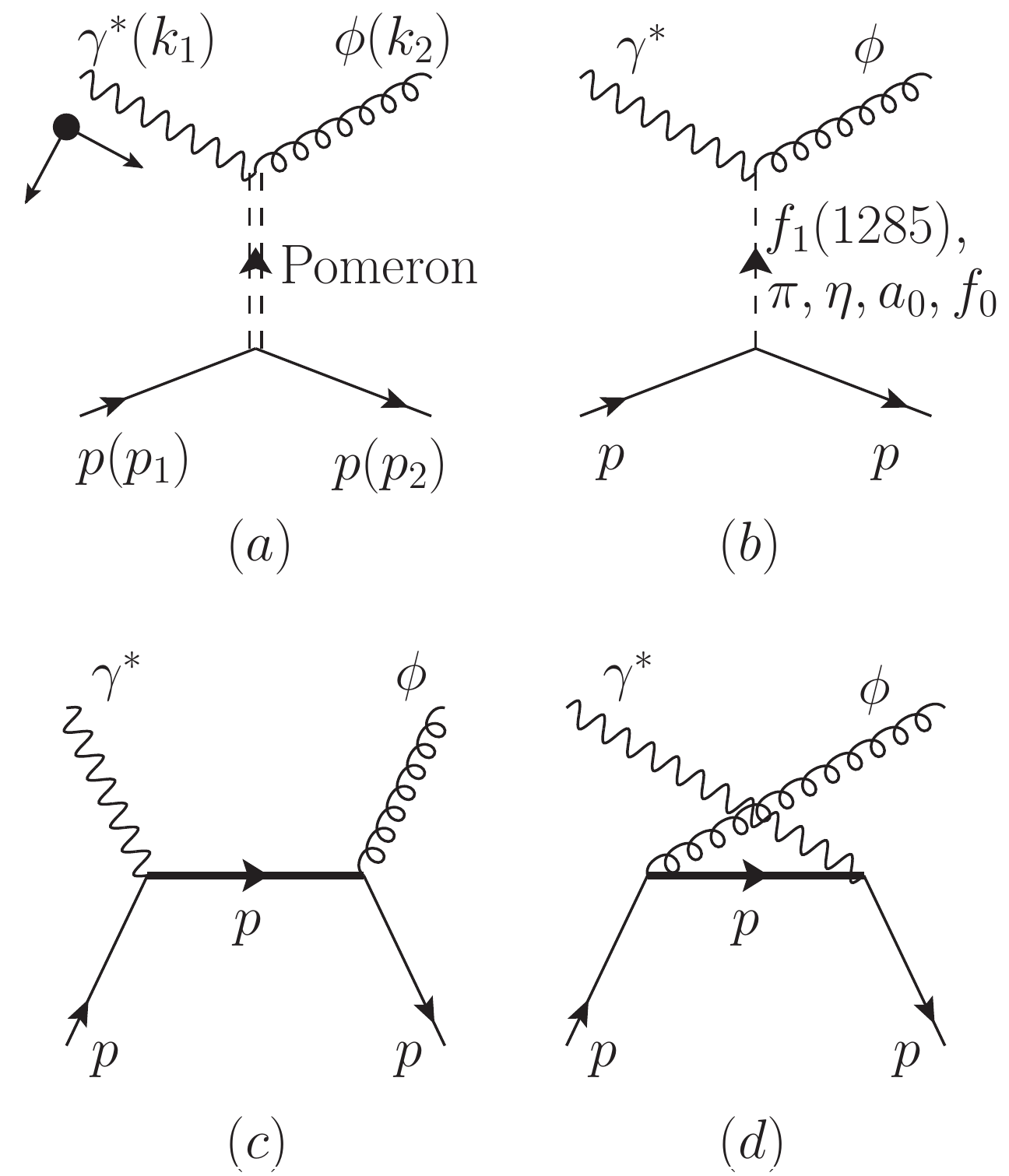}
\caption{Tree-level Feynman diagrams for $\gamma^* p\to \phi p$, which include
Pomeron ($a$), Reggeized $f_1(1285)$ axial-vector-meson, ($\pi$,$\eta$)
pseudoscalar-meson, and ($a_0$,$f_0$) scalar-meson exchanges in the $t$ channel
($b$), and direct $\phi$-meson radiations via the proton in the $s$ and $u$
channels ($c$ and $d$).}
\label{FIG2:FD}
\end{figure}

We employ an effective Lagrangian approach here.
The production mechanisms under consideration are drawn with the relevant
Feynman diagrams in Fig.~\ref{FIG2:FD},
which includes Pomeron ($\mathbb P$) ($a$), Reggeized $f_1(1285)$ axial-vector
(AV)-meson, ($\pi$,$\eta$) pseudoscalar (PS)-meson, and ($a_0$,$f_0$) scalar
(S)-meson exchanges in the $t$ channel ($b$), and direct $\phi$-meson radiations
via the proton in the $s$ and $u$ channels ($c$ and $d$).

We can write the invariant amplitude as
\begin{align}
\label{eq:AmpDef}
\mathcal{M} =
\varepsilon_\nu^*(\lambda) \bar{u}_{N'}(\lambda_f)
\mathcal{M}^{\mu\nu} u_N(\lambda_i) \epsilon_\mu(\lambda_\gamma),
\end{align}
the helicities of the particles being given in parentheses.
The forms of the invariant amplitudes are in complete analogy to the $\phi$-meson
photoproduction case~\cite{Kim:2019kef} for the corresponding diagrams.
The Dirac spinors of the incoming and outgoing nucleons are designated by $u_N$
and $u_{N'}$, respectively.
$\epsilon_\mu$ and $\varepsilon_\nu$ denote the polarization vectors of the
virtual photon and the $\phi$ meson, respectively.
The extension of the photo- to electro-production of mesons entails an
additional longitudinal component ($\lambda_\gamma=0$) for the virtual-photon
polarization vector in addition to the transverse ($\lambda_\gamma=\pm 1$) ones:
\begin{align}
\label{eq:PhoPolVec}
\epsilon (\pm 1) = \frac{1}{\sqrt{2}} (0,\,\mp 1,\,-i,\,0),\,\,\,
\epsilon (0) = \frac{1}{\sqrt{Q^2}} (k,\,0,\,0,\, E_{\gamma^*}),
\end{align}
where $E_{\gamma^*} = \sqrt{k^2-Q^2} = (M_N\nu-Q^2)/W$.
The polarization vectors of the virtual photon and the $\phi$ meson satisfy the
conventional completeness relations~\cite{Obukhovsky:2009th}
\begin{align}
\label{eq:ComplRela}
\sum_{\lambda_\gamma = 0,\pm 1} (-1)^{\lambda_\gamma}
\epsilon_\mu(\lambda_\gamma) \epsilon_\nu^*(\lambda_\gamma) =
g_{\mu\nu} - \frac{k_{1\mu}k_{1\nu}}{k_1^2}, \,\,\,                        
\sum_{\lambda = 0,\pm 1} \varepsilon_\mu(\lambda)
\varepsilon_\nu^*(\lambda) = - \left[ g_{\mu\nu} -
\frac{k_{2\mu}k_{2\nu}}{M_\phi^2} \right].
\end{align}

\subsection{Pomeron exchange}
\begin{figure}[h]
\includegraphics[width=4.5cm]{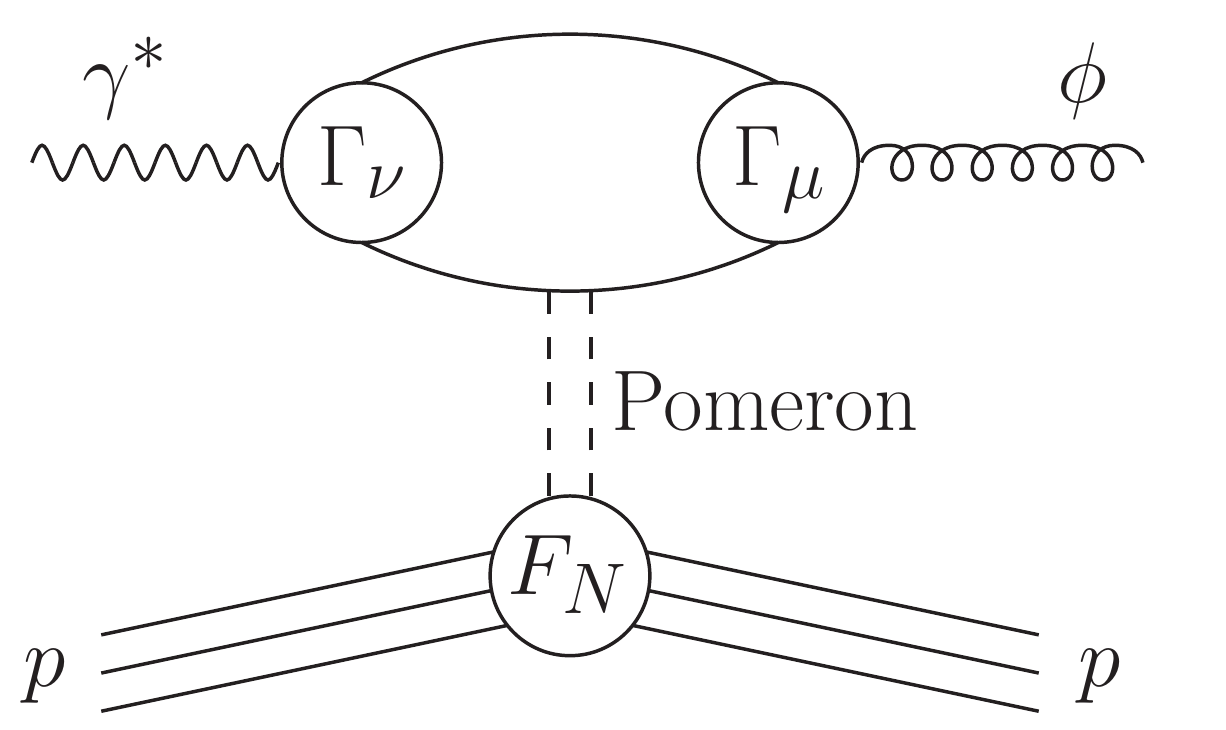}
\caption{Quark diagram for Pomeron exchange in the DL model, based on the
Pomeron-isoscalar-photon analogy.}
\label{FIG3:QD}
\end{figure}

Figure~\ref{FIG2:FD}(a) draws the Pomeron exchange that governs the scattering
process in the high-energy and small $t$ regions.
We follow the Donnachie-Landshoff (DL) model~\cite{Donnachie:1987abc} where a
microscopic description of the Pomeron exchange in vector meson photo- and
electroproduction is given in terms of nonperturbative Reggeized-two-gluon
exchange based on the Pomeron-isoscalar-photon analogy
(see Fig.~\ref{FIG3:QD})~\cite{Pichowsky:1996jx,Pichowsky:1996tn}.

As a consequence, the invariant amplitude for the Pomeron exchange can be
expressed as
\begin{align}
\label{eq:AmpPom}
\mathcal{M}_{\mathbb P}^{\mu\nu} = - M_{\mathbb P}(s,t)
\left[
\left( g^{\mu\nu} - \frac{k_2^\mu k_2^\nu}{k_2^2} \right) \rlap{\,/}{k_1} -
\left( k_1^\nu - \frac{k_2^\nu k_1 \cdot k_2}{k_2^2} \right) \gamma^\mu -
\left( \gamma^\nu - \frac{\rlap{\,/}{k_2}k_2^\nu}{k_2^2} \right) k_2^\mu \right].
\end{align}
One find that the last term in the square bracket on the right-hand side (r.h.s.)
breaks the gauge invariance and the following modification is performed to make
it gauge invariant~\cite{Titov:2003bk}:
\begin{align}
\label{eq:k2Mod}
k_2^\mu \to k_2^\mu - \frac{(p_1+p_2)^\mu k_1 \cdot k_2}{(p_1+p_2)
\cdot k_1}.
\end{align}
Other prescriptions for the spin structure for conserving the gauge invariance
are detailed in Ref.~\cite{Titov:1998bw} but are not used in this work, because
qualitative descriptions of $\phi$-meson electroproduction with them are found
to be very poor.
The scalar function in Eq.~(\ref{eq:AmpPom}) is given by
\begin{align}
\label{eq:ScaFun}
M_{\mathbb P}(s,t) = C_{\mathbb P} F_\phi(t) F_N(t) \frac{1}{s}
\left( \frac{s}{s_{\mathbb P}} \right)^{\alpha_{\mathbb P}(t)}
\mathrm{exp} \left[ -\frac{i\pi}{2}\alpha_{\mathbb P}(t) \right].
\end{align}
The strength factor is determined to be $C_{\mathbb P}=3.6$ and the energy-scale
factor to be $s_{\mathbb P}=(M_N+M_\phi)^2$.
$F_N(t)$ and $F_\phi(t)$ stand for the nucleon isoscalar electromagnetic (EM)
form factor~\cite{Donnachie:1988nj,Laget:1994ba} and the form factor of the
$\gamma {\mathbb P} \phi$ coupling~\cite{Jaroszkiewicz:1974ep,Donnachie:1983hf},
respectively, and take the forms
\begin{align}
\label{eq:FF:Pom}
F_N(t) = \frac{4M_N^2- 2.8t}{(4M_N^2-t)(1-t/0.71)^2},\,\,\,\,\,
F_\phi(t)=\frac{2 \mu_0^2 \Lambda_\phi^2}{(Q^2+\Lambda_\phi^2-t)
(2\mu_0^2+ Q^2+\Lambda_\phi^2-t)},
\end{align}
The mass scale $\Lambda_\phi$ is proportional to the quark mass of the loop
diagram in Fig.~\ref{FIG3:QD} and is chosen to be $\Lambda_\phi^2 = M_\phi^2$
as done previously.
The momentum scale is given by $\mu_0^2 = 1.1\, \mathrm{GeV}^2$ and the Pomeron
trajectory is known to be $\alpha_\mathbb{P}(t)=1.08+0.25t$.

\subsection{$f_1(1285)$ axial-vector meson exchange}
We first consider Reggeized $f_1(1285)$ AV meson among meson exchanges depicted
in Fig.~\ref{FIG2:FD}(b).
Its importance is indicated in elastic $p$-$p$ scattering and elastic
photoproduction of $\rho$ and $\phi$ mesons due to its special relation to the
axial anomaly through the matrix elements of the flavor singlet axial vector
current~\cite{Kochelev:1999zf} and is confirmed more specifically in
$\phi$-meson photoproduction done by the authors~\cite{Kim:2019kef}.
Thus we include $f_1(1285)$ AV-meson exchange in $\phi$-meson electroproduction.

The effective Lagrangian for the $AVV$ vertex is obtained from the hidden gauge
approach~\cite{Kaiser:1990yf}
\begin{align}
\label{eq:Lag:GPhif1}
\mathcal{L}_{\gamma \phi f_1} = g_{\gamma \phi f_1}
\epsilon^{\mu\nu\alpha\beta} \partial_\mu A_\nu
\partial^\lambda \partial_\lambda \phi_\alpha f_{1\beta} ,
\end{align}
where $f_1$ denotes the $f_1(1285)$ field with its quantum number $I^G(J^{PC})=
0^+(1^{++})$.
The experimental data for the branching ratio (Br)
${\mathrm{Br}}_{f_1 \to \phi \gamma} = 7.5 \times 10^{-4}$ and the decay width
$\Gamma_{f_1}$ = 22.7 MeV~\cite{Tanabashi:2018oca} lead to
\begin{align}
\label{eq:CC:GPhif1}
g_{\gamma \phi f_1} = 0.17 \, \mathrm{GeV^{-2}},
\end{align}
from Eq.~(\ref{eq:Lag:GPhif1}).

The effective Lagrangian of the AV meson interaction with the nucleon takes the
form
\begin{align}
\label{eq:Lag:f1NN}
\mathcal{L}_{f_1 NN} = - g_{f_1 NN} \bar N
\left[ \gamma_\mu - i \frac{\kappa_{f_1 NN}}{2M_N} \gamma_\nu \gamma_\mu
\partial^\nu \right] f_1^\mu \gamma_5 N .
\end{align}
The coupling constant $g_{f_1NN}$ is obtained to be~\cite{Birkel:1995ct}
\begin{align}
\label{eq:CC:f1NN}
g_{f_1NN} = 2.5 \pm 0.5,
\end{align}
and we use the maximum value $g_{f_1NN} = 3.0$.
Although the tensor term can have an effect on $\phi$-meson electroproduction,
we take the value of $\kappa_{f_1 NN}$ to be zero in this work for brevity.

The corresponding invariant amplitude reads
\begin{align}
\label{eq:Amp:M}
\mathcal{M}_{f_1}^{\mu\nu} = i \frac{M_\phi^2 g_{\gamma \phi f_1}g_{f_1 NN}}{t-M_{f_1}^2}
\epsilon^{\mu\nu\alpha\beta}
\left[ -g_{\alpha\lambda}+\frac{q_{t\alpha} q_{t\lambda}}{M_{f_1}^2} \right]
\left[ \gamma^\lambda + \frac{\kappa_{f_1 NN}}{2M_N} \gamma^\sigma \gamma^\lambda
q_{t\sigma} \right] \gamma_5 k_{1 \beta},
\end{align}
where $q_t=k_2-k_1$.
We substitute the exchange of the entire $f_1(1285)$ Regge trajectory for the
above single $f_1(1285)$ meson exchange as~\cite{Donachie2002}
\begin{eqnarray}
\label{eq:REGGEPRO}
P_{f_1}^{\mathrm{Feyn}}(t) = \frac{1}{t-M_{f_1}^2} \to
P_{f_1}^{\mathrm{Regge}}(t) = \left(\frac{s}{s_{f_1}} \right)^{\alpha_{f_1}(t)-1}
\frac{\pi\alpha'_{f_1}}{\sin[\pi\alpha_{f_1}(t)]}
\frac{1}{\Gamma [\alpha_{f_1}(t)]} D_{f_1}(t),
\end{eqnarray}
such that the spin structures of the interaction vertices are kept and the
Regge propagator effectively interpolates between small- and large-momentum
transfers and can contribute to the high energy region properly.
The Regge trajectory is determined to be $\alpha_{f_1}(t) = 0.99 + 0.028t$
~\cite{Kochelev:1999zf} and the energy-scale factor $s_{f_1} = 1\,\mathrm{GeV}^2$.
The odd signature factor is given by~\cite{Kochelev:1999zf}
\begin{eqnarray}
\label{eq:SigFac}
D_{f_1}(t) = \frac{-1 + \mathrm{exp}(-i\pi\alpha_{f_1}(t))}{2}.
\end{eqnarray}

The invariant amplitude is modified by introducing the following form factors
\begin{align}
\label{eq:FF:GVf1:f1NN:1}
g_{\gamma \phi f_1} \to g_{\gamma \phi f_1} F_{\gamma \phi f_1} (Q^2,t),\,\,\,
g_{f_1 NN} \to g_{f_1 NN} F_{f_1 NN} (t),
\end{align}
where
\begin{align}
\label{eq:FF:GVf1:f1NN:2}
F_{\gamma \phi f_1} (t,Q^2) = \frac{\Lambda_{f_1}^2 - M_{f_1}^2}{\Lambda_{f_1}^2 - t}
\frac{\Lambda_q^2}{\Lambda_q^2+Q^2},\,\,\,
F_{f_1 NN} (t) = \frac{\Lambda_{f_1}^2 - M_{f_1}^2}{\Lambda_{f_1} ^2 - t},
\end{align}
which are normalized at $t=M_{f_1}^2$ and $Q^2=0$ as $F_{\gamma \phi f_1}
= F_{f_1 NN} = 1$.

\subsection{Pseudoscalar- and scalar-meson exchanges}
Figure~\ref{FIG2:FD}(b) also includes the contributions of the $t$-channel
($\pi$,$\eta$) PS- and ($a_0$,$f_0$) S-meson exchange diagrams.
The EM interaction Lagrangians for the PS- and S-meson exchanges,
respectively, can be written as
\begin{align}
\label{eq:Lag:GMPhi}
\mathcal{L}_{\gamma\Phi\phi} =& \frac{eg_{\gamma\Phi\phi}}{M_\phi}
\epsilon^{\mu\nu\alpha\beta} \partial_\mu A_\nu \partial_\alpha \phi_\beta \Phi ,\cr
\mathcal{L}_{\gamma S \phi} =& \frac{eg_{\gamma S \phi}}{M_\phi}
F^{\mu\nu} \phi_{\mu\nu} S,
\end{align}
where $\Phi=\pi^0(135,0^-),\,\eta(548,0^-)$ and $S=a_0(980,0^+),\,f_0(980,0^+)$.
The field-strength tensors for the photon and $\phi$-meson  are given by
$F^{\mu\nu} = \partial^\mu A^\nu - \partial^\nu A^\mu$ and $\phi^{\mu\nu} =
\partial^\mu \phi^\nu - \partial^\nu \phi^\mu$, respectively, and $e$ the unit
electric charge.
The EM coupling constants are deduced from the $\phi \to \Phi \gamma$ and $\phi
\to S \gamma$ decay widths
\begin{align}
\label{eq:Wid:GMPhi}
\Gamma_{\phi \to \Phi \gamma} =
\frac{e^2}{12\pi} \frac{g_{\gamma\Phi\phi}^2}{M_\phi^2}
\left( \frac{M_\phi^2- M_\Phi^2}{2M_\phi} \right)^3,\,\,\,
\Gamma_{\phi \to S \gamma} =
\frac{e^2}{3\pi} \frac{g_{\gamma S \phi}^2}{M_\phi^2}
\left( \frac{M_\phi^2- M_S^2}{2M_\phi} \right)^3.
\end{align}
With the $\phi$-meson branching ratios of
${\mathrm{Br}}_{\phi \to \pi \gamma} = 1.30 \times 10^{-3}$,
${\mathrm{Br}}_{\phi \to \eta \gamma} = 1.303 \times 10^{-2}$,
${\mathrm{Br}}_{\phi \to a_0 \gamma} = 7.6 \times 10^{-5}$, and
${\mathrm{Br}}_{\phi \to f_0 \gamma} = 3.22 \times 10^{-4}$
and the value of $\Gamma_\phi = 4.249$ MeV~\cite{Tanabashi:2018oca}, we obtain
\begin{align}
\label{eq:CC:GMPhi}
g_{\gamma\pi\phi}=-0.14, \hspace{1em} g_{\gamma\eta\phi}=-0.71, \hspace{1em}
g_{\gamma a_0 \phi}= -0.77, \hspace{1em} g_{\gamma f_0 \phi}= -2.44.
\end{align}

The strong interaction Lagrangians for the PS- and S-meson exchanges read
\begin{align}
\label{eq:Lag:MNN}
\mathcal{L}_{\Phi NN} =& -ig_{\Phi NN} \bar N \Phi \gamma_5 N ,                \cr
\mathcal{L}_{S NN} =& -g_{S NN} \bar N SN ,
\end{align}
respectively.
We use the following strong coupling constants determined by the Nijmegen
potential~\cite{Stoks:1999bz,Rijken:1998yy}:
\begin{align}
\label{eq:CC:MNN}
g_{\pi NN}= 13.0, \hspace{1em} g_{\eta NN}= 6.34,   \hspace{1em}
g_{a_0 NN}= 4.95, \hspace{1em} g_{f_0 NN}= -0.51.
\end{align}

We obtain the invariant amplitudes for PS- and S-meson exchanges as
\begin{align}
\label{eq:Amp:PSS}
\mathcal{M}_\Phi^{\mu\nu} =& i\frac{e}{M_\phi}
\frac{g_{\gamma\Phi\phi}g_{\Phi NN}}{t-M_\Phi^2} \epsilon^{\mu\nu\alpha\beta} k_{1\alpha}
k_{2\beta} \gamma_5 ,
\cr
\mathcal{M}_S^{\mu\nu} =&-\frac{e}{M_\phi}
\frac{2g_{\gamma S \phi}g_{S NN}}{t- M_S^2 + i\Gamma_S M_S}
(k_1 \cdot k_2 g^{\mu\nu} - k_2^\mu k_1^\nu),
\end{align}
respectively, where we use $M_{a_0} = 980$ MeV, $M_{f_0} = 990$ MeV, and
$\Gamma_{a_0,f_0} = 75$ MeV~\cite{Tanabashi:2018oca}.

Here we also consider the form factors $F_{\gamma M \phi} (t,Q^2)$ and $ F_{MNN} (t)$
for each vertex describing the dependence on the $t$ and $Q^2$ similar to
Eqs.~(\ref{eq:FF:GVf1:f1NN:1}) and (\ref{eq:FF:GVf1:f1NN:2}):
\begin{align}
\label{eq:FF:GMV:MNN:2}
F_{\gamma M \phi} (t,Q^2) = \frac{\Lambda_M^2 - M_M^2}{\Lambda_M^2 - t}
\frac{\Lambda_q^2}{\Lambda_q^2+Q^2},\,\,\,
F_{M NN} (t) = \frac{\Lambda_M^2 - M_M^2}{\Lambda_M ^2 - t},
\end{align}
where $M=(\Phi,S)$.

\subsection{Direct $\phi$-meson radiation term}
It is argued that the direct $\phi$-meson radiation term drawn in
Figs.~\ref{FIG2:FD}(c) and~\ref{FIG2:FD}(d) gives a small contribution to the
unpolarized cross sections but a very distinct contribution to some polarization
observables in $\phi$-meson photoproduction~\cite{Titov:1999eu,Kim:2019kef}.
Thus it is interesting to include this term in $\phi$-meson electroproduction.

The effective Lagrangians for the direct $\phi$-meson radiation contributions
can be written as
\begin{align}
\label{eq:LAG:N}
\mathcal{L}_{\gamma NN} =& - e \bar N
\left[ \gamma_\mu - \frac{\kappa_N}{2M_N} \sigma_{\mu\nu} \partial^\nu
\right] N A^\mu,                                                          \cr
\mathcal{L}_{\phi NN} =& - g_{\phi NN} \bar N
\left[ \gamma_\mu - \frac{\kappa_{\phi NN}}{2M_N} \sigma_{\mu\nu} \partial^\nu
\right] N \phi^\mu,
\end{align}
where the anomalous magnetic moment of the proton is $\kappa_p = 1.79$
~\cite{Tanabashi:2018oca} and the vector and tensor coupling constants for the
$\phi$-meson to the nucleon are determined to be $g_{\phi NN}=-0.24$ and
$\kappa_{\phi NN}=0.2$ ~\cite{Meissner:1997qt}.

The $\phi$-radiation invariant amplitudes are computed as
\begin{align}
\label{eq:Amp:N}
\mathcal{M}_{\phi\,\mathrm{rad},s}^{\mu\nu} &= \frac{e g_{\phi NN}}{s-M_N^2}
\left(\gamma^\nu - i\frac{\kappa_{\phi NN}}{2M_N}
\sigma^{\nu\alpha} k_{2\alpha} \right) (\rlap{/}{q}_s + M_N)
\left( \gamma^\mu{F_1^p} + i{F_2^p}\frac{\kappa_N}{2M_N}
\sigma^{\mu\beta} k_{1\beta} \right),
\cr
\mathcal{M}_{\phi\,\mathrm{rad},u}^{\mu\nu} &= \frac{e g_{\phi NN}}{u-M_N^2}
\left( \gamma^\mu{F_1^p} + i{F_2^p}\frac{\kappa_N}{2M_N}
\sigma^{\mu\alpha} k_{1\alpha} \right) (\rlap{/}{q}_u + M_N)
\left(\gamma^\nu - i\frac{\kappa_{\phi NN}}{2M_N}
\sigma^{\nu\beta} k_{2\beta}
\right),
\end{align}
for the $s$ and $u$ channels, respectively, with the EM form factors being
involved.
$q_{s,u}$ are the four momenta of the exchanged particles, i.e., $q_s=k_1+p_1$ and
$q_u=p_2-k_1$.

Note that the Ward-Takahashi identity (WTI) is violated when a different form
for the form factor $F_1^p$ is used for the electric terms of the two invariant
amplitudes.
Thus we use the same form and can check the sum of them restores the WTI as
\begin{align}
\label{eq:Amp:N:GI}
\mathcal{M}_{\phi\,\mathrm{rad},s}^{\mathrm{elec}} (\epsilon \to k_1)&=
\frac{e g_{\phi NN}}{2k_1 \cdot p_1 - Q^2}
\bar{u}_{N'} \left( \rlap{/}{\varepsilon^*} + \frac{\kappa_{\phi NN}}{4M_N}
(\rlap{/}{\varepsilon^*} \rlap{/}{k}_2 - \rlap{/}{k}_2 \rlap{/}{\varepsilon^*})
\right) (2k_1 \cdot p_1 - Q^2) {F_1^p} u_N,
\cr
\mathcal{M}_{\phi\,\mathrm{rad},u}^{\mathrm{elec}} (\epsilon \to k_1)&=
\frac{- e g_{\phi NN}}{2k_1 \cdot p_2 + Q^2} \bar{u}_{N'}
{F_1^p} (2k_1 \cdot p_2 + Q^2)
\left( \rlap{/}{\varepsilon^*} + \frac{\kappa_{\phi NN}}{4M_N}
(\rlap{/}{\varepsilon^*} \rlap{/}{k}_2 - \rlap{/}{k}_2 \rlap{/}{\varepsilon^*})
\right) u_N,
\end{align}
such that $\mathcal{M}_{\phi\,\mathrm{rad}}^{\mathrm{elec}}(\epsilon \to k_1) \propto
(F_1^p - F_1^p)=0$.
We follow the suggestion given by David and Workman~\cite{Davidson:2001rk} for
the form factors:
\begin{align}
\label{eq:Amp:N:Simple}
\mathcal{M}_{\phi\,\mathrm{rad}} =
(\mathcal{M}_{\phi\,\mathrm{rad},s}^{\mathrm{elec}} +
\mathcal{M}_{\phi\,\mathrm{rad},u}^{\mathrm{elec}})
F_c(s,u)^2 {F_1^p(Q^2)}
+ \mathcal{M}_{\phi\,\mathrm{rad},s}^{\mathrm{mag}} F_N(s)^2 {F_2^p(Q^2)}
+ \mathcal{M}_{\phi\,\mathrm{rad},u}^{\mathrm{mag}} F_N(u)^2 {F_2^p(Q^2)}.
\end{align}
Here a common form factor is introduced which conserves the on-shell condition
and the crossing symmetry:
\begin{align}
\label{eq:FF:C}
F_c (s,u) = 1 - [1-F_N(s)][1-F_N(u)],
\end{align}
with
\begin{align}
\label{eq:FF:N}
F_N (x) = \frac{\Lambda^4_N}{\Lambda^4_N+(x-M^2_N)^2}, \,\,\,x=(s,u).
\end{align}
Since the magnetic terms are self-gauge-invariant, the form of
Eq.~(\ref{eq:FF:N}) is just used for them.

Now, we give some details for the Dirac ($F_1$) and Pauli ($F_2$) form
factors by using their relations with the Sachs one ($G_{E,M}$)
~\cite{Nam:2017yeg}:
\begin{align}
\label{eq:G:EM}
G_E(Q^2) = F_1(Q^2) - \kappa_N \tau F_2(Q^2),\,\,\,
G_M(Q^2) = \mu_N G_E(Q^2) = F_1(Q^2) + \kappa_N F_2(Q^2),
\end{align}
where $\tau = Q^2/4M_p^2$ and correspondingly,
\begin{align}
F_1(Q^2) = \frac{G_E(Q^2) + \tau G_M(Q^2)}{1+\tau}, \,\,\,
F_2(Q^2) = \frac{G_M(Q^2) - G_E(Q^2)}{\kappa_N(1+\tau)}.
\end{align}
The Sachs form factors are parametrized for the proton and the neutron in the
literature by
\begin{align}
G_E^p(Q^2) \simeq G_D(Q^2),\,\,G_M^p(Q^2) \simeq \mu_p G_D(Q^2),\,\,
G_E^n(Q^2) \simeq - \frac{a\mu_n\tau}{1+b\tau} G_D(Q^2),\,\,
G_M^n(Q^2) \simeq \mu_n G_D(Q^2),
\end{align}
with the dipole-type of form factor
\begin{align}
\label{eq:DipoleFF}
G_D(Q^2) = \left[ \frac{1}{1+Q^2 \langle r^2 \rangle_E^p/12} \right]^2 ,
\end{align}
where the electric root-mean-squared charge radius of the proton is given
by (0.863 $\pm$ 0.004) fm~\cite{Kelly:2004hm}.

\section{Numerical results and Discussions}
\label{SecIV}
We now discuss our numerical results from the present work.
The remaining model parameters are the cutoff masses involved in the form
factors.
The cutoff masses for the $t$ dependent form factors for meson exchanges are
determined to be $\Lambda_{f_1,a_0,f_0}=1.4$ and $\Lambda_{\pi,\eta}= 0.6$ GeV and
that for the $\phi$-meson radiations in Eq.~(\ref{eq:FF:N}) to be
$\Lambda_N = 1.0$ GeV.
We have shown that those phenomenological form factors provide a good
description of $\phi$-meson photoproduction at the considered energy region
$W$ = (2$-$3) GeV and at even much higher one $W \lesssim 10$ GeV as well
in our recent work~\cite{Kim:2019kef}.
The cutoff masses for the $Q^2$ dependent form factors
[$\Lambda_q^2/(\Lambda_q^2+Q^2)$] for all meson exchanges
are chosen to be $\Lambda_q = 0.9$ GeV in common.

The cross section dependence on the angle $\Phi$ of meson electroproduction is
decomposed into the transverse (T), longitudinal (L), and interference (TT, LT)
parts as
\begin{eqnarray}
\label{dsdPhi}
\frac{d\sigma}{d\Phi} = \frac{1}{2\pi}
(\sigma + \varepsilon \sigma_{\mathrm{TT}} \cos2 \Phi +
\sqrt{2\varepsilon(1+\varepsilon)} \sigma_{\mathrm{LT}} \cos\Phi),
\end{eqnarray}
where $\sigma = \sigma_{\mathrm{T}} + \varepsilon \sigma_{\mathrm{L}}$.
We refer to Appendix~\ref{AppenAAA} for the explicit expressions for the T-L
separated differential cross sections.
If helicity is conserved in the $s$ channel (SCHC), then the second and
third terms vanish.
The virtual-photon polarization parameter $\varepsilon$ is defined by
\begin{align}
\label{eq:def:StrLongPol}
\varepsilon = \left[ 1 + \frac{2k^2}{Q^2} \mathrm{tan}^2 \frac{\theta_e}{2}
\right]^{-1}.
\end{align}
In all our calculations, we fix it to be $\varepsilon =0.5$ because the
available data to be used are carried out with the value close to it.

\begin{figure}[htp]
\centering
\includegraphics[width=7cm]{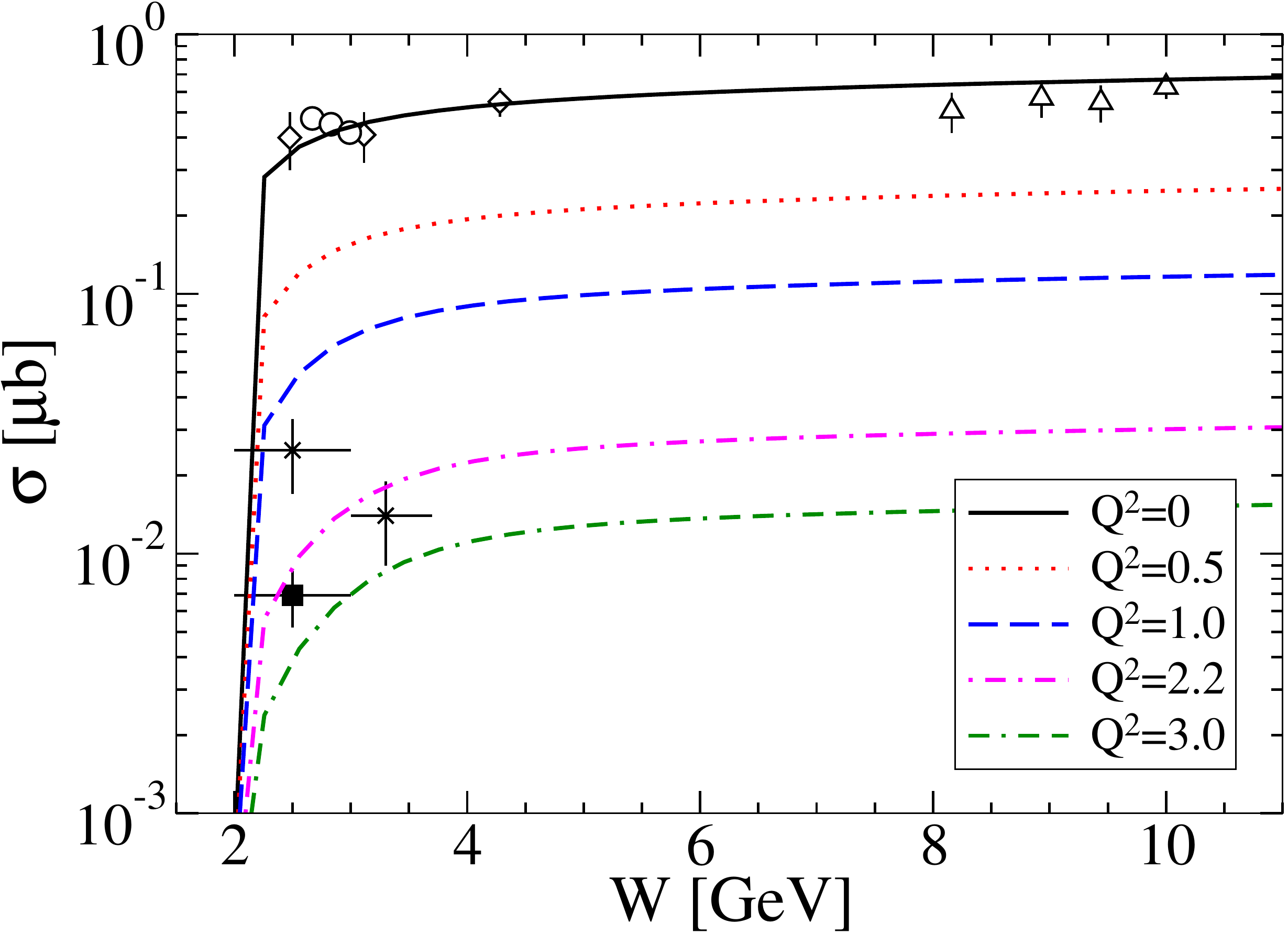}
\caption{Total cross sections for $\gamma^* p \to \phi p$ are plotted as a
function of $W$ for five different photon virtualities
$Q^2=(0,\,0.5,\,1.0,\,2.2,\,3.0)\,\mathrm{GeV}^2$.
The $\phi$ photoproduction ($Q^2=0$) data are from Refs.~\cite{Ballam:1972eq}
(diamond),~\cite{Barber:1981fj} (circle), and ~\cite{Egloff:1979mg} (triangle).
The Cornell~\cite{Cassel:1981sx} (star) and CLAS~\cite{Santoro:2008ai} (square)
data correspond to the results at $Q^2$ = 2.2 $\mathrm{GeV}^2$.}
\label{fig:4}
\end{figure}
Figure~\ref{fig:4} displays the results of the total cross sections as a
function of $W$ for five different photon virtualities $Q^2$.
The slowly rising total cross sections with increasing $W$ are kept for all
values of $Q^2$ due to the dominant Pomeron contribution.
The agreement with the experimental data~\cite{Ballam:1972eq,Barber:1981fj,
Egloff:1979mg} is good at the real photon limit $Q^2=0$ over the whole energy
range.
The magnitude of the total cross section when $Q^2=0.5\,\mathrm{GeV}^2$ reaches
the level around $40\%$ relative to that when $Q^2=0$.
The results get more suppressed for higher values of $Q^2$.

\begin{figure}[htp]
\centering
\includegraphics[width=12.4cm]{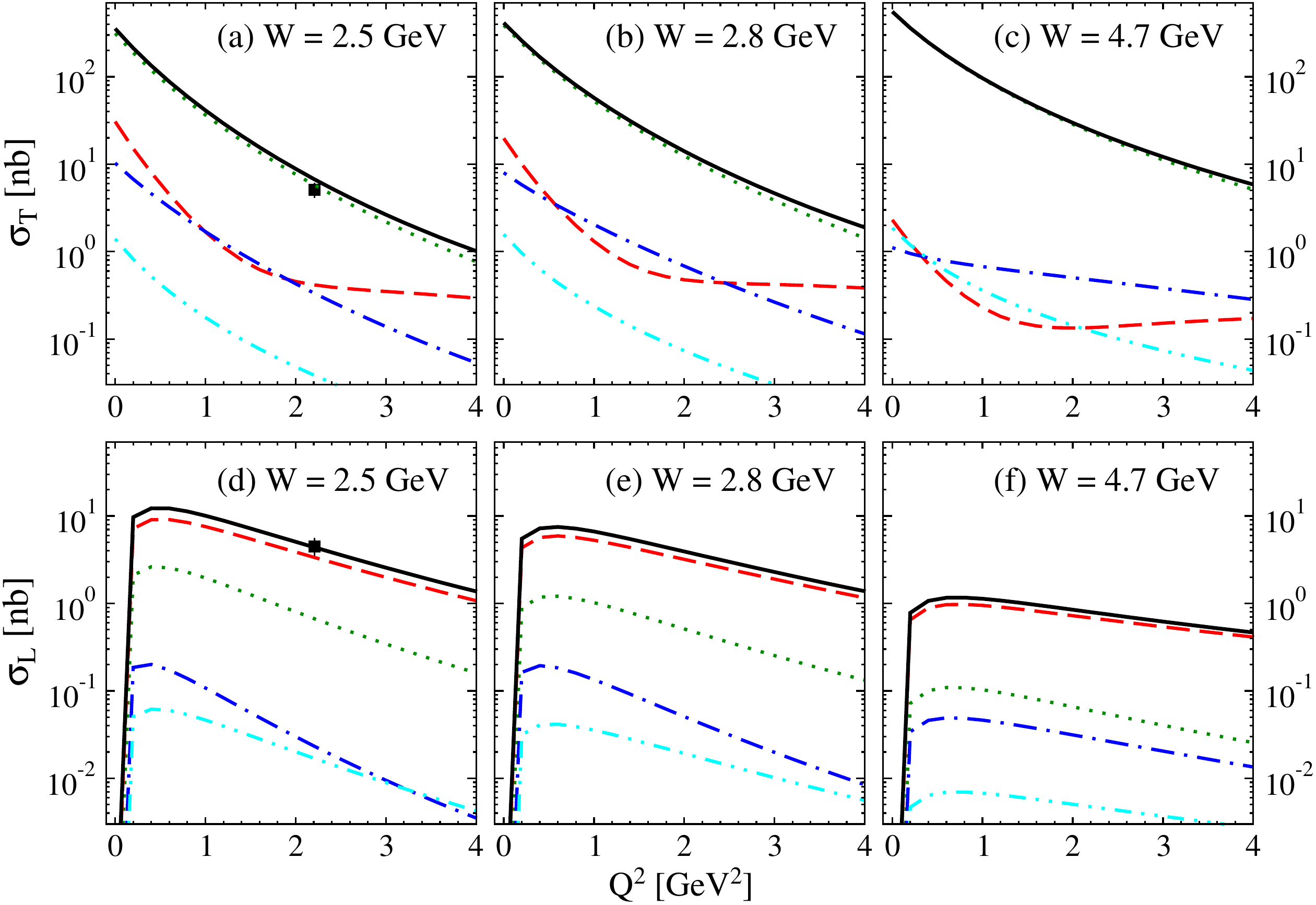}
\caption{Transverse $\sigma_{\mathrm{T}}$ [(a)-(c)] and longitudinal
$\sigma_{\mathrm{L}}$ [(d)-(f)] cross sections are plotted as functions of $Q^2$ for
three different c.m. energies labeled on each subplot.
The green dotted, cyan dot-dot-dashed, blue dot-dashed, and red dashed curves
stand for the contributions from the individual Pomeron, AV-meson, PS-meson, and
S-meson exchanges, respectively.
The black solid curves indicate the total contribution.
The CLAS data in panels (a) and (d) are extracted from
Ref.~\cite{Santoro:2008ai}.}
\label{fig:5}
\end{figure}
It is essential to investigate the separated components of the cross sections to
clarify the different role of considered meson exchanges.
The CLAS Collaboration~\cite{Santoro:2008ai} extracted the interference cross
sections to be $\sigma_{\mathrm{TT}} = -1.1 \pm 3.1$ nb and $\sigma_{\mathrm{LT}} =
2.2 \pm 1.1$ nb from six bins of the $d\sigma/d\Phi$ data measured in the range
of $W$ = (2.0$-$3.0) $\mathrm{GeV}$ and $Q^2$ = (1.4$-$3.8) $\mathrm{GeV^2}$
 using the relation Eq.~(\ref{dsdPhi}).
The matrix element is extracted as well to be $r_{00}^{04}=0.33
\pm 0.12$ from five bins of the polar angular distribution $W (\cos\theta_H)$.
With the additional assumption of SCHC, the ratio of the longitudinal to
transverse cross section is obtained to be
$R = \sigma_{\mathrm L}/\sigma_{\mathrm T} = 1.05 \pm 0.38$.
Also $r_{1-1}^1= 0.38 \pm 0.23$ and $R = 0.72 \pm 0.3$ are obtained from eight
bins of the angular distribution $W (\psi=\phi_H-\Phi)$ under the SCHC
approximation.
Lastly, the longitudinal cross section is calculated to be
$\sigma_{\mathrm L} (Q^2=2.21\,\mathrm{GeV}^2) = 4.5 \pm 1.1$ nb using the values
of the average ratio $R=0.85 \pm 0.24$ and of the average cross section
$\sigma = 6.9 \pm 1.7$ nb~\cite{Santoro:2008ai}.
The definitions of the matrix elements will be given later
~\cite{Schilling:1973ag}. 

Figure~\ref{fig:5} depicts the results of the transverse ($\sigma_{\mathrm{T}}$)
and longitudinal ($\sigma_{\mathrm{L}}$) cross sections as functions of $Q^2$ in
the upper and lower panels, respectively, for three different c.m. energies $W$.
We find that a predominant mechanism that contributes to the transverse cross
section $\sigma_{\mathrm{T}}$ is the Pomeron exchange.
The individual AV-, PS-, and S-meson exchanges all have little influences on
$\sigma_{\mathrm T}$ for three $W = (2.5,\,2.8,\,4.7)$ GeV energy values.
The contribution of the Pomeron exchange at $W$ = 2.5 GeV and $Q^2$ = 2.21
$\mathrm{GeV}^2$ is in very good agreement with the CLAS data as shown in
Fig.~\ref{fig:5}(a).
However, it is quite the opposite in the case of $\rho$-meson electroproduction.
That is, the PS-meson exchange governs the transverse cross section
$\sigma_{\mathrm T}$
due to the $M1$ spin transition $\gamma_{\mathrm T}^* + \pi^0(\eta) \to \rho^0$
and Pomeron exchange is relatively much more suppressed at low c.m. energies
($W\sim \mathrm{a\,few}\,\,\mathrm{GeV}$) and low photon virtualities ($Q^2 \sim
\mathrm{a\,few}\,\,\mathrm{GeV}^2$) ~\cite{Obukhovsky:2009th}.
In this work, employing a strong form factor for the PS-meson exchange obviously
overestimates the available $\phi$-meson electroproduction CLAS data.
Thus we use a rather small value of the cutoff mass as $\Lambda_{\pi,\eta}=0.6$
GeV.

Meanwhile, the contribution of the Pomeron exchange alone for the longitudinal
cross section $\sigma_{\mathrm{L}}$ is an order of magnitude smaller than that for
$\sigma_{\mathrm{T}}$ at $W$ = 2.5 GeV and $Q^2$ = (1$-$4) $\mathrm{GeV}^2$ as
shown in Fig.~\ref{fig:5}(d).
The difference becomes larger for higher c.m. energies $W=2.8,\,4.7$ GeV.
The inclusion of S-meson exchange to the Pomeron exchange gives a sufficiently
better description of $\sigma_{\mathrm{L}}$.
That is how the S-meson exchange form factor is determined, i.e.,
$\Lambda_{a_0,f_0}= 1.4$ GeV.
As displayed in Fig.~\ref{fig:5}(d)-\ref{fig:5}(f),
the contribution of the S-meson exchange
for $\sigma_{\mathrm{L}}$ is highly enhanced relative to $\sigma_{\mathrm{T}}$ and
even prevails over that of the Pomeron exchange over all the ranges of $W$ and
$Q^2$.
All the amplitudes for the longitudinal photons must vanish at the limit
$Q^2=0$ and this behavior is imposed explicitly in our calculation.
For both $\sigma_{\mathrm{T}}$ and $\sigma_{\mathrm{L}}$ cases, the contribution
of the AV-meson exchange is comparable to that of the PS-meson exchange or even
more suppressed.

\begin{figure}[tp]
\centering
\includegraphics[width=12cm]{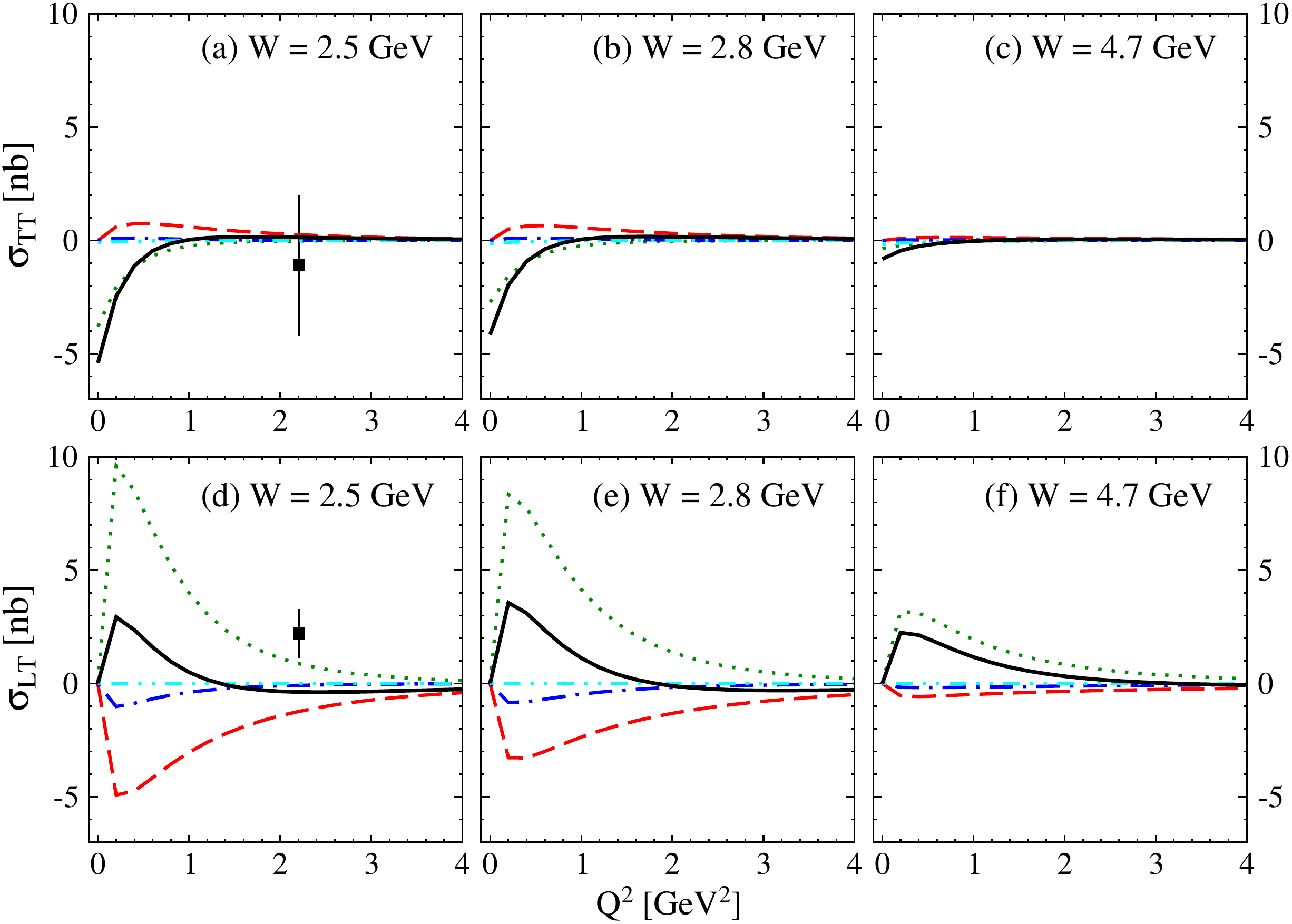}
\caption{The same as in Fig.~\ref{fig:5} but for the interference cross sections
$\sigma_{\mathrm{TT}}$ (upper panels) and $\sigma_{\mathrm{LT}}$ (lower panels),
respectively.}
\label{fig:6}
\end{figure}
We present the results of the interference cross sections $\sigma_{\mathrm{TT}}$
and $\sigma_{\mathrm{LT}}$ in the upper and lower panels in Fig.~\ref{fig:6},
respectively.
The overall results are the decrease of the absolute magnitudes with increasing
$W$ for all contributions, indicative of SCHC at relatively higher values of $W$
and $Q^2$.
The meson-exchange contributions are all close to zero for $\sigma_{\mathrm{TT}}$
and the total contribution is entirely dependent on the Pomeron exchange which
is the strongest at $Q^2=0$ and gradually decreases with increasing $Q^2$.
At $Q^2 \geq 1\,\mathrm{GeV}^2$, the results are consistent with zero for all
values of $W$, which are within the CLAS data at $W$ = 2.5 GeV and
$Q^2$ = 2.21 $\mathrm{GeV}^2$ as shown in Fig.~\ref{fig:6}(a).

Different patterns are observed for the results of the individual
$\sigma_{\mathrm{LT}}$ cross sections in comparison to $\sigma_{\mathrm{TT}}$ as seen
in Fig.~\ref{fig:6}(d)-\ref{fig:6}(f).
The overall positive sign applies to the Pomeron contribution for
$\sigma_{\mathrm{LT}}$.
It is peaked at about 0.3 $\mathrm{GeV}^2$ and falls off with
increasing $Q^2$.
The signs of the PS- and S-meson contributions are the same each other but
are opposite to that of the Pomeron contribution.
The CLAS data shown in Fig.~\ref{fig:6}(d) is close to the Pomeron contribution.
However, the inclusion of PS- and S-meson exchanges pulls down
$\sigma_{\mathrm{LT}}$ and finally the total contribution reaches zero at
$Q^2$ = 2.21 $\mathrm{GeV}^2$.
That is one more reason why PS-meson exchange should be suppressed in
$\phi$-meson electroproduction.
The small increase of the cutoff masses for both the PS- and S-meson form factors
from the present ones makes the total results of $\sigma_{\mathrm{LT}}$ worse.
The AV-meson exchange contributes almost negligibly to both $\sigma_{\mathrm{TT}}$
and $\sigma_{\mathrm{LT}}$.
However, note that its contribution to $\sigma_{\mathrm{TT}}$ is much more
sensitive than that to $\sigma_{\mathrm{LT}}$ under the variation of the cutoff
mass $\Lambda_{f_1}$.

\begin{figure}[htp]
\centering
\includegraphics[width=11cm]{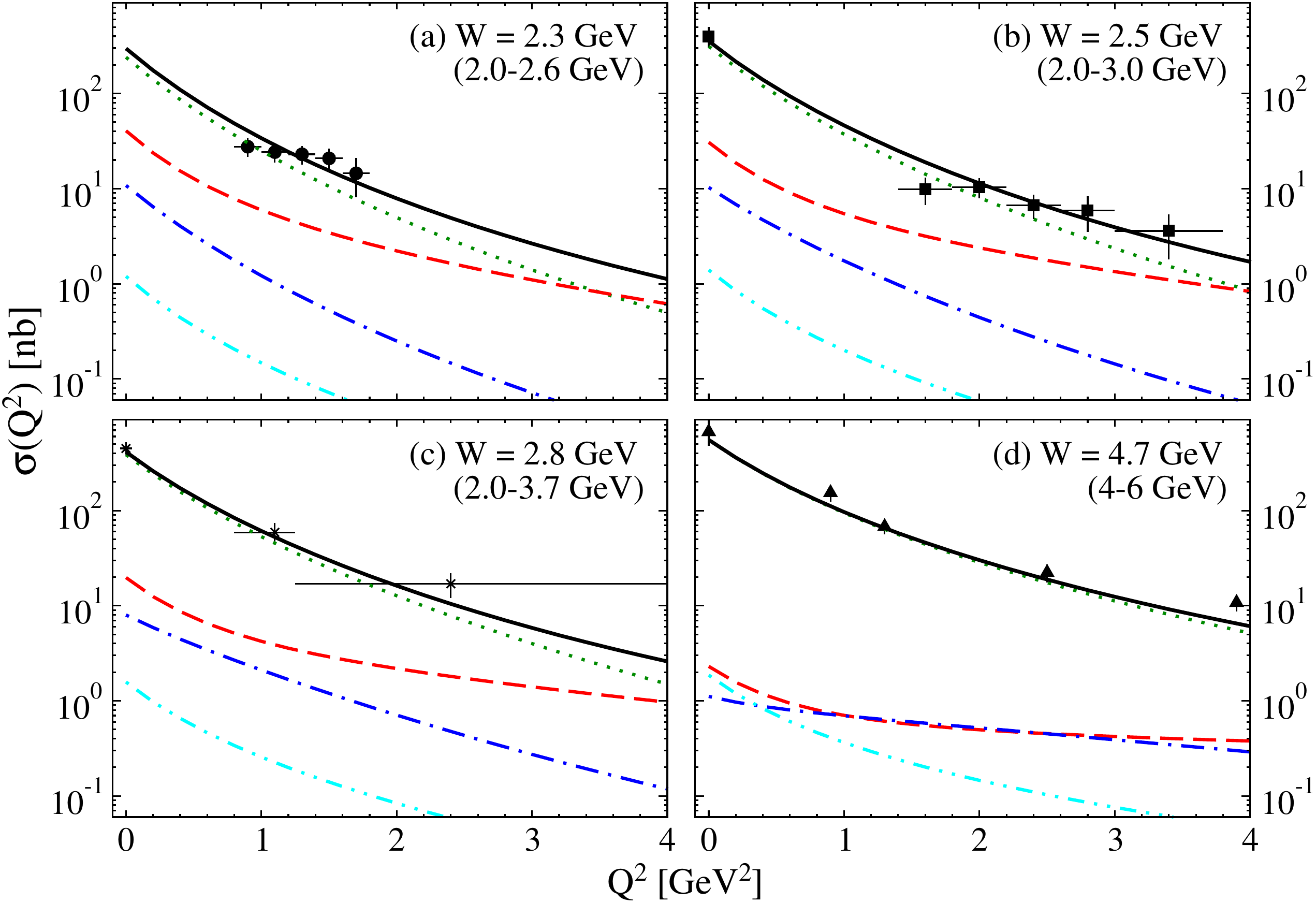}
\caption{Total cross sections are plotted as functions of $Q^2$ for four
different c.m. energies labeled on each subplot.
The curves are defined in the caption of Fig.~\ref{fig:5}.
The circle~\cite{Lukashin:2001sh} and square~\cite{Santoro:2008ai} data are
from the CLAS Collaboration.
The star and triangle data from the Cornell~\cite{Cassel:1981sx} and
HERMES Collaboration~\cite{Borissov:2000wb}, respectively.
(b),(c),(d) The photoproduction points at $Q^2=0$ are from Refs.
~\cite{Ballam:1972eq,Barber:1981fj}.}
\label{fig:7}
\end{figure}
\begin{figure}[htp]
\centering
\includegraphics[width=12cm]{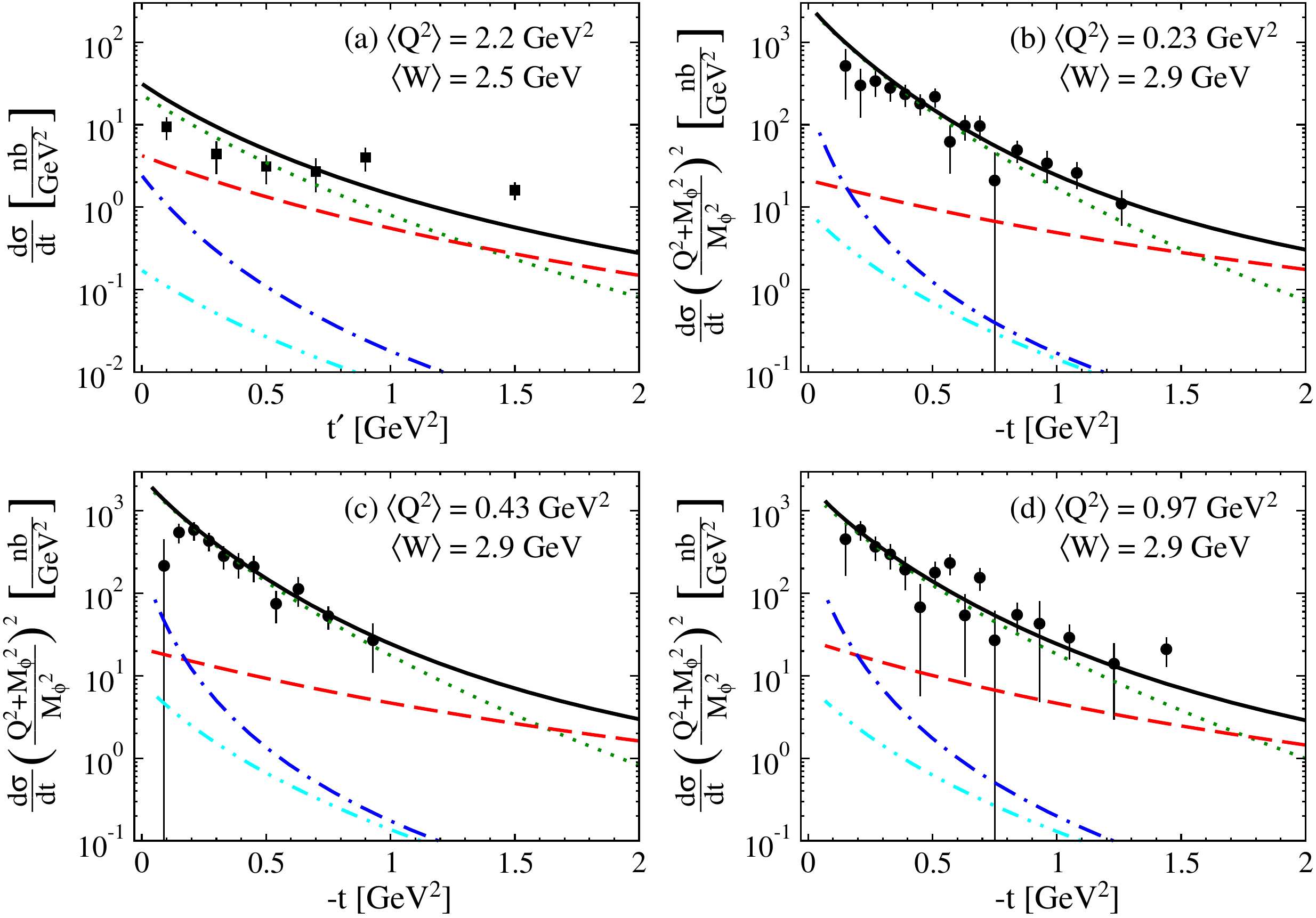}
\caption{(a) Differential cross section $d\sigma/dt$ is plotted as a function of
$t^\prime = |t-t_{min}|$ and compared with the CLAS data~\cite{Santoro:2008ai}.
(b),(c),(d) $d\sigma/dt$ multiplied by the factor $[(Q^2+M_\phi^2)/M_\phi^2]^2$ is
plotted as functions of $-t$ for three different photon virtualities labeled on
each subplot at $W=2.9$ GeV and compared with the Cornell
data~\cite{Dixon:1977pg,Dixon:1978vy}.
The curves are defined in the caption of Fig.~\ref{fig:5}.}
\label{fig:8}
\end{figure}

Figure~\ref{fig:7} depicts the results of the unpolarized total cross sections
as functions of $Q^2$ for four different c.m. energies $W$.
The model parameters are all constrained previously from the study of the
separated cross sections although the available data on a Rosenbluth
separation~\cite{Rosenbluth:1950yq} are too poor to be a reliable basis for
verifying the $\phi$-meson electroproduction mechanism.
It is interesting that the unpolarized cross sections are also well described
over the whole kinematical ranges of $W$ and $Q^2$, since the Pomeron exchange is
mainly responsible for describing the experimental data.
Although small, the effects of meson exchanges are revealed at larger values of
$Q^2$ mostly due to the milder slope of the S-meson contribution than
the Pomeron one.
Our effective hadronic model accounts for the points at the real photon limit
$Q^2=0$ as expected from Fig.~\ref{fig:4},

The results of the differential cross section $d\sigma/dt$ are displayed in
Fig.~\ref{fig:8}(a) at $Q^2$ = 2.2 $\mathrm{GeV}^2$ and $W$ = 2.5 GeV as a
function of $t'\equiv |t-t_{\mathrm{min}}|$ where $t_{\mathrm{min}}$ stands for the
minimum value of $t$ at fixed values $Q^2$ and $W$.
Figures~\ref{fig:8}(b)-\ref{fig:8}(d) depict the results of $d\sigma/dt$, which
are multiplied by the factor $[(Q^2+M_\phi^2)/M_\phi^2]^2$ to eliminate the
$\phi$-propagator dependence, as functions of $-t$ for the three different
photon virtualities $Q^2$ at $W=2.9$ GeV.
They corroborate our finding that the dominant contribution is the Pomeron
exchange by which the $t$ dependence is properly described.
The strength of the S-meson exchange becomes larger than that of the
Pomeron exchange at $|t| \gtrsim 1.5\,\mathrm{GeV}^2$.
It should be mentioned that the $a_0$- and $\pi$-meson contributions are more
important than those of the $f_0$ and $\eta$ mesons, respectively.

\begin{figure}[htp]
\centering
\includegraphics[width=7cm]{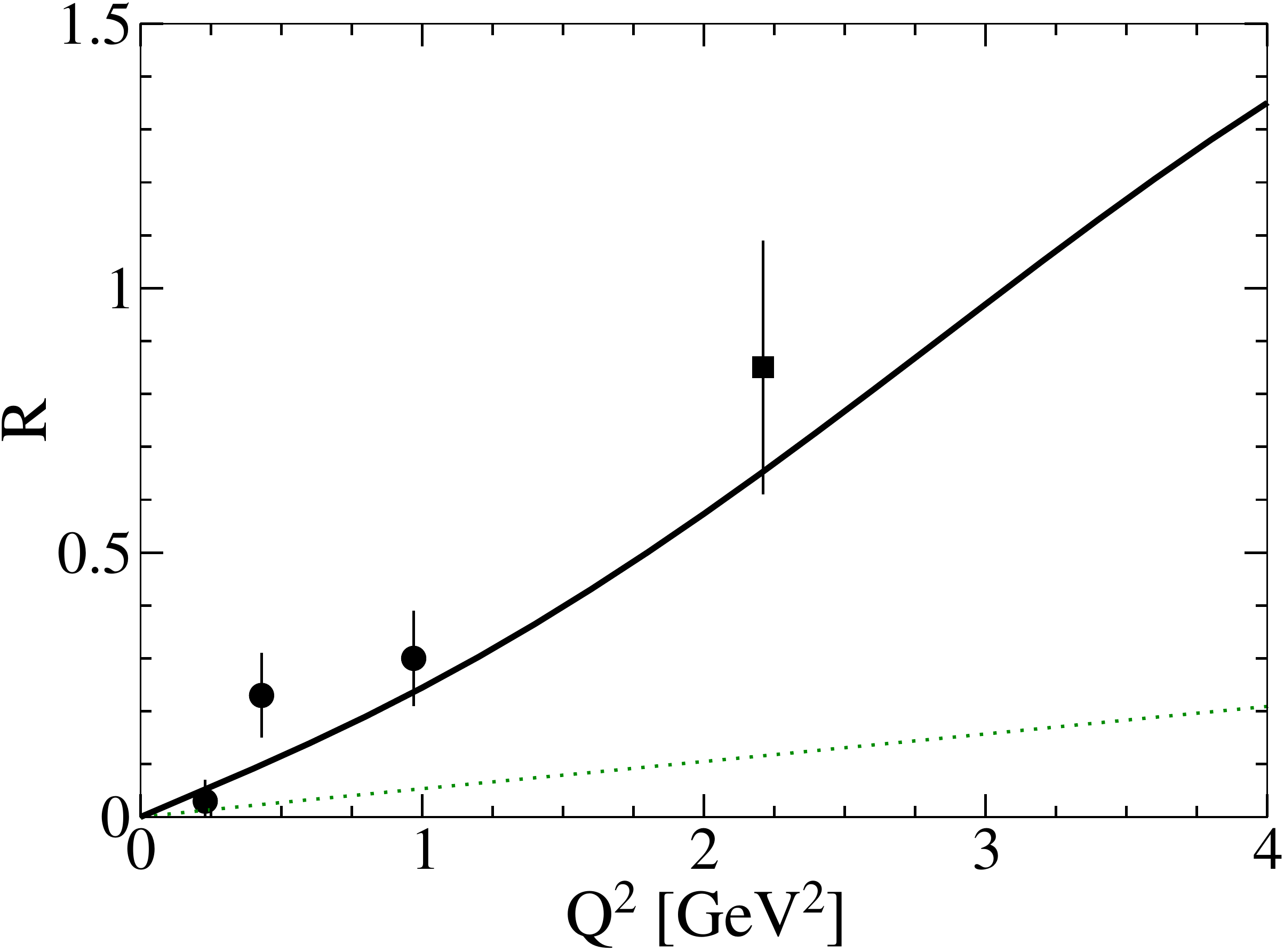}
\caption{The ratio of cross sections for longitudinally and transversely
polarized photons $R=\sigma_{\mathrm L}/\sigma_{\mathrm T}$ is plotted as a function
of $Q^2$ at $W$ = 2.5 GeV.
The green dotted and black solid curves stand for the Pomeron and total
contributions, respectively.
The data are from the Cornell~\cite{Dixon:1978vy} (circle) and
CLAS Collaboration~\cite{Santoro:2008ai} (square).}
\label{fig:9}
\end{figure}
\begin{figure}[htp]
\centering
\includegraphics[width=12cm]{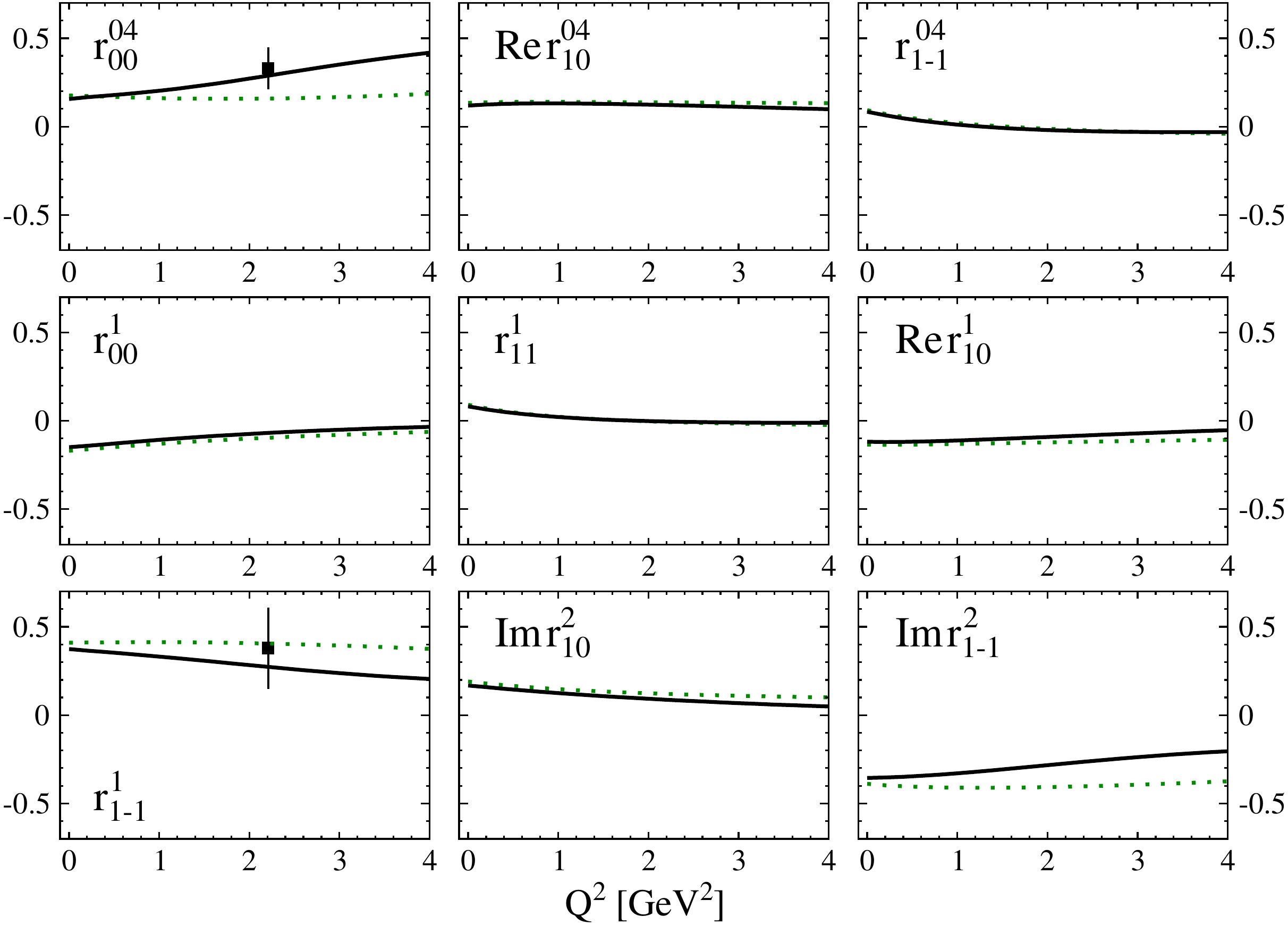}
\caption{Various matrix elements are plotted as functions of $Q^2$ at $W$ = 2.5
GeV.
The extracted data for $r_{00}^{04}$ and $r_{1-1}^1$ are from the CLAS
Collaboration~\cite{Santoro:2008ai}.}
\label{fig:10}
\end{figure}

We present the results of the ratio of the longitudinal to transverse cross
section $R=\sigma_{\mathrm L}/\sigma_{\mathrm T}$ in Fig.~\ref{fig:9} at $W$ = 2.5
GeV although the Cornell data (circle) correspond to $W$ = 2.9 GeV.
Our results yield $R \leq 0.2$ for the Pomeron contribution, meaning that it
mainly involves the transverse component in our given ranges of $Q^2$.
The agreement is noticeably better when the S-meson exchange is additionally
included.
Note that as the fixed c.m. energy $W$ increases, the rise of $R$ with respect
to $Q^2$ becomes smaller as indicated in Fig.~\ref{fig:5} where the transverse
and longitudinal components of the total contribution exhibit the opposite
pattern with $W$.

Finally, we make our predictions of various matrix elements, denoted by
$r_{ij}^\alpha$ and related to the $\phi$-meson spin-density matrix elements
(SDMEs)~\cite{Schilling:1973ag}.
When the experiments cannot conduct a $\sigma_{\mathrm L}/\sigma_{\mathrm T}$
separation, $r_{ij}^\alpha$ can be represented as
\begin{align}
\label{eq:def:UnsepME}
r_{ij}^{04} &= \frac{\rho_{ij}^0+\varepsilon R \rho_{ij}^4}{1+\varepsilon R}, \cr
r_{ij}^\alpha &= \frac{\rho_{ij}^\alpha}{1+\varepsilon R},
\hspace{1em}\mathrm{for}\hspace{0.5em}\alpha = (0-3),                     \cr
r_{ij}^\alpha &= \sqrt{R}\frac{\rho_{ij}^\alpha}{1+\varepsilon R},
\hspace{1em}\mathrm{for}\hspace{0.5em}\alpha = (5-8 ).
\end{align}
We refer to Appendix~\ref{AppenBBB} for the definitions of the SDMEs
$\rho_{ij}^\alpha$.
The matrix elements are described in the helicity frame, where the $\phi$ meson
is at rest and its quantization axis is chosen to be antiparallel to the momentum
of the outgoing proton in the c.m. frame of the hadron production process.

The results are shown in Fig.~\ref{fig:10} as functions of $Q^2$ for nine
different matrix elements labeled on each subplot.
If SCHC holds, all presented matrix elements becomes zero except for
$r_{00}^{04}$, $r_{1-1}^1$, and $\mathrm{Im}\,r_{1-1}^2$.
It turns out that the SCHC approximation indeed applies at
$Q^2$ = (1$-$4) $\mathrm{GeV}^2$.
This conclusion is consistent with the results of Fig.~\ref{fig:6} where SCHC is
verified from the interference cross sections $\sigma_{\mathrm{TT}}$ and
$\sigma_{\mathrm{LT}}$.
A good agreement with the CLAS data is obtained for $r_{00}^{04}$ and $r_{1-1}^1$.

It is useful to examine the relative contribution between the natural ($N $)
and unnatural ($U$) parity exchange processes for the transverse cross section.
The parity asymmetry is defined by~\cite{Donachie2002,Schilling:1973ag}
\begin{align}
\label{eq:def:ParityAsy}
P \equiv \frac{\sigma_T^N-\sigma_T^U}{\sigma_T^N+\sigma_T^U} =
(1 + \varepsilon R)(2r_{1-1}^1 - r_{00}^1),
\end{align}
and our results yield $P \simeq$ 0.85 and 0.95 for the Pomeron and total
contributions, respectively, at $W$ = 2.5 GeV and $Q^2$ = (0$-$4)
$\mathrm{GeV^2}$.
This observation indicates that the dominant mechanism of the transverse cross
section is the natural-parity exchange.
Also when a purely natural-parity exchange is considered, e.g., the S-meson
exchange, we have the following relation~\cite{Tytgat,Morand:2005ex}:
\begin{align}
\label{eq:def:NPE}
1 - r_{00}^{04} + 2r_{1-1}^{04} - 2r_{11}^1 - 2r_{1-1}^1 = 0.
\end{align}
For the Pomeron contribution, our results yield the values close to zero.
The results from the total contribution yield $\simeq$ 0.1 and thus the
dominance of the natural-parity exchange is again verified.

\section{Summary}
\label{SecV}
We have investigated the reaction mechanism of $\phi$-meson electroproduction
off the proton target based on the gauge-invariant effective Lagrangians in the
tree-level Born approximation.
Each contribution of the Pomeron, Reggeized $f_1(1285)$ AV-meson, ($\pi$,$\eta$)
PS-meson, and ($a_0$,$f_0$) S-meson exchanges is scrutinized in the $t$ channel
diagram employing the CLAS and Cornell data.
The direct $\phi$-meson radiations via the proton are also taken into account
in the $s$ and $u$ channels simultaneously to conserve gauge invariance.
We summarize the essential points on which our model calculations have performed.

\begin{itemize}
\item The unpolarized cross sections $\sigma$ show slow rising with increasing
$W$.
The results for the real photon limit $Q^2=0$ match with the available data very
well.
The magnitudes of $\sigma$ become smaller with increasing $Q^2$ and the main
contribution turns out to be Pomeron exchange.
The cross sections $\sigma$ give us no insight into what meson exchanges
contributes to $\phi$-meson electroproduction and thus it is necessary to
examine the T-L separated cross sections.

\item The Pomeron and S-meson exchanges dominate the transverse
($\sigma_{\mathrm T}$) and longitudinal ($\sigma_{\mathrm L}$) cross sections,
respectively, at $Q^2$ = (0$-$4) $\mathrm{GeV}^2$ for three considered
$W = 2.5,\,2.8,\,4.7$ GeV photon energies.
Pomeron exchange alone is sufficient to describe the CLAS data on
$\sigma_{\mathrm T}$ at $W$ = 2.5 GeV and $Q^2$ = 2.21 $\mathrm{GeV}^2$.
Other meson contributions are more suppressed and the difference between the
Pomeron and meson contributions becomes larger as $W$ increases.
Also the CLAS data on $\sigma_{\mathrm L}$ at the same $W$ and $Q^2$ values are
accounted for solely by the S-meson exchange.
The Pomeron contribution for $\sigma_{\mathrm T}$ falls off faster than the
S-meson contribution for $\sigma_{\mathrm L}$ as $Q^2$ increases.

\item The interference cross sections $\sigma_{\mathrm{TT}}$ and
$\sigma_{\mathrm {LT}}$ are found to be useful to clarify the role of AV- and
PS-meson exchanges which is difficult from the study of $\sigma_{\mathrm T}$
and $\sigma_{\mathrm L}$.
First, for $\sigma_{\mathrm {TT}}$, all meson-exchange contributions are
compatible to zero and the total contribution is determined entirely by the
Pomeron exchange which is negative in sign and the strongest at $Q^2=0$ and
falls off steadily with increasing $Q^2$.
However, note that the AV-meson exchange is relatively more sensitive than other
meson exchanges under the variation of each cutoff mass and its role can be more
clearly verified when more experimental data on $\sigma_{\mathrm{TT}}$ are produced.
Next, for the case of $\sigma_{\mathrm{LT}}$, the patterns of each contribution are
totally different from $\sigma_{\mathrm{TT}}$.
The Pomeron contribution is peaked at about  $Q^2 = 0.3\,\mathrm{GeV}^2$
and falls off gradually with increasing $Q^2$.
The S-meson and PS-meson contributions also show similar peak positions.
The signs of these two contributions are the same each other but are opposite to
that of the Pomeron contribution.
If the S-meson and PS-meson contributions increase from the present ones, then
the total contribution will become negative and deviate from the CLAS data at
$W$ = 2.5 GeV and $Q^2$ = 2.21 $\mathrm{GeV}^2$.
Thus the S-meson and PS-meson contributions must be small from the present
results.
The effect of the AV-meson exchange is almost negligible.
For both $\sigma_{\mathrm{TT}}$ and $\sigma_{\mathrm {LT}}$ cases, the total
contribution is close to zero at $Q^2$ = (1$-$4) $\mathrm{GeV}^2$, indicating
SCHC.

\item The $t$ dependence of the differential cross sections $d\sigma/dt$ is
also well described in the range of $|t|$ = (0$-$2) $\mathrm{GeV}^2$.
The Pomeron exchange dominates over the whole $t$ regions.
Although small, the role of the meson exchanges are significant only at
$|t| \gtrsim 1\,\mathrm{GeV}^2$ mostly from the S-meson exchange.
The ratio of the longitudinal to transverse cross section
$R = \sigma_{\mathrm L}/\sigma_{\mathrm T}$ rises linearly with increasing $Q^2$.
$R$ for the total contribution is about 7 times larger than that for the Pomeron
contribution due to the S-meson contribution and is good agreement with the
experimental data.

\item We examine various matrix elements defined in helicity frame which is in
favor of SCHC.
We find that the values of $\mathrm{Re}\,r_{10}^{04}$, $r_{1-1}^{04}$, $r_{00}^1$,
$r_{11}^1$, $\mathrm{Re}\,r_{10}^1$, and $\mathrm{Im}\,r_{10}^2$ are close to zero
at $Q^2$ = ($1-$4) $\mathrm{GeV}^2$ and thus SCHC holds.
Our results match with the CLAS data on $r_{00}^{04}$ and $r_{1-1}^1$.
When we come to $\rho$-meson electroproduction, in the similar low $Q^2$ and $W$
ranges, it is difficult to draw a firm conclusion concerning SCHC although most
physical observables seem to support SCHC~\cite{Morrow:2008ek}.
On the contrary, SCHC is known to be broken in $w$-meson electroproduction
because of the different contributions of Pomeron and various meson
exchanges~\cite{Morand:2005ex}.
Also a small but significant violation of SCHC is found in $\phi$-meson
electroproduction in the high ranges of $Q^2$ and $W$~\cite{Adloff:2000nx},
where the generalized parton distributions (GPD) and factorization of scales will
become relevant.

\item The parity asymmetry provides us with the information of the relative
strength of the natural to unnatural parity exchanges in the $t$ channel for
$\sigma_{\mathrm T}$.
Our results yield $P \simeq$ 0.95, implying that the transverse cross section is
mainly governed by the natural-parity exchange.
Meanwhile, the contribution of the direct $\phi$-meson radiations via the proton
is found to be almost negligible.
\end{itemize}
The currently available data on $\phi$-meson electroproduction are very limited
and new experiments at current or future electron facilities are strongly called
for.
We can gain a deeper understanding of $\phi$-meson electroproduction mechanism
by comparing our numerical results with those of the GPD-based model.
It is valuable to extend the present work to electroproductions of $\rho$,
$\omega$, and $J/\psi$ mesons. The corresponding work is underway.

\section*{ACKNOWLEDGMENTS}
The authors are grateful to A.~Hosaka (RCNP) for fruitful discussions.
This work is supported in part by the National Research Foundation of Korea
(NRF) funded by the Ministry of Education, Science and Technology (MSIT)
(NRF-2018R1A5A1025563). The work of S.H.K. was supported by NRF-2019R1C1C1005790.
The work of S.i.N. was also supported partly by NRF-2019R1A2C1005697.
\appendix
\section{APPENDIX A:
T-L SEPARATED DIFFERENTIAL CROSS SECTIONS}
\label{AppenAAA}
The separated components of the differential cross sections in the Rosenbluth
formula~\cite{Rosenbluth:1950yq} take the forms
\begin{align}
\label{eq:SepaAmpl}
\frac{1}{\mathcal{N}} \frac{d\sigma_{\mathrm{T}}}{dt} & =
\frac12 \sum_{\lambda_\gamma=\pm 1}
\overline{|{\mathcal M}^{(\lambda_\gamma)}|^2},                                \cr
\frac{1}{\mathcal{N}} \frac{d\sigma_{\mathrm{L}}}{dt} & =
\overline{|{\mathcal M}^{(\lambda_\gamma=0)}|^2},                              \cr
\frac{1}{\mathcal{N}} \frac{d\sigma_{\mathrm{TT}}}{dt} & =
-\frac12 \sum_{\lambda_\gamma=\pm 1}
\overline{{\mathcal M}^{(\lambda_\gamma)} {\mathcal M}^{{(-\lambda_\gamma)}^*}},   \cr
\frac{1}{\mathcal{N}} \frac{d\sigma_{\mathrm{LT}}}{dt} & =
-\frac{1}{2\sqrt{2}} \sum_{\lambda_\gamma=\pm 1} \lambda_\gamma
(\overline{{\mathcal M}^{(0)} {\mathcal M}^{{(\lambda_\gamma)}^*}} +
\overline{{\mathcal M}^{(\lambda_\gamma)} {\mathcal M}^{(0)^*}}),
\end{align}
for vector-meson electroproduction.
The common kinematical factor $\mathcal{N}$ is defined by
\begin{align}
\label{eq:StanConst}
{\mathcal N} = [32\pi(W^2-M_N^2)Wk]^{-1}.
\end{align}
Here the squared invariant amplitude is expressed as
\begin{align}
\label{eq:AveSumAmpl}
\overline{|{\mathcal M}^{(\lambda_\gamma)}|^2}
= \frac12 \sum_{\lambda_i,\lambda_f,\lambda}
{\mathcal M}_{\lambda_f\lambda;\lambda_i\lambda_\gamma}
{\mathcal M}_{\lambda_f\lambda;\lambda_i\lambda_\gamma}^*,
\end{align}
where the averaging over the incoming nucleon ($\lambda_i$) helicity and the
summation over the outgoing $\phi$ meson ($\lambda$) and nucleon
($\lambda_f$) helicities are indicated.
${\mathcal M}_{\lambda_f\lambda;\lambda_i\lambda_\gamma}^*$ stands for the complex
conjugate of the amplitude
${\mathcal M}_{\lambda_f\lambda;\lambda_i\lambda_\gamma}$.
Note that the differential cross sections have the following relations:
\begin{eqnarray}
\label{eq:DCS:relation}
\frac{d\Omega_\phi}{dt} = \frac{\pi}{|k||p|}.
\end{eqnarray}

\section{APPENDIX B: SPIN-DENSITY MATRIX ELEMENTS}
\label{AppenBBB}
We obtain nine components of the spin-density matrix elements (SDMEs) of the
$\phi$ meson if those of the virtual photon are decomposed into the standard set
of nine matrices $\Sigma^\alpha
(\alpha = 0-8)$~\cite{Schilling:1973ag}:
\begin{align}
\label{eq:def:SDME}
\rho_{\lambda \lambda'}^\alpha &= \frac{1}{2N_\alpha}
\sum_{\lambda_\gamma, \lambda_\gamma', \lambda_i, \lambda_f}
{\mathcal M}_{\lambda_f \lambda ; \lambda_i \lambda_\gamma}
\Sigma_{\lambda_\gamma \lambda_\gamma'}^\alpha
{\mathcal M}_{\lambda_f \lambda'; \lambda_i \lambda_\gamma'}^* ,
\end{align}
where the normalization factors $N_\alpha$ are defined by
\begin{align}
\label{eq:def:NorFac}
N_\alpha &= N_{\mathrm T} =\frac12
\sum_{\lambda_\gamma=\pm 1, \lambda, \lambda_i, \lambda_f}
|{\mathcal M}_{\lambda_f \lambda ; \lambda_i \lambda_\gamma}|^2
\hspace{1em}\mathrm{for}\hspace{0.5em}\alpha = (0-3),                     \cr
N_\alpha &= N_{\mathrm L} = \sum_{\lambda, \lambda_i, \lambda_f}
|{\mathcal M}_{\lambda_f \lambda ; \lambda_i 0}|^2,
\hspace{1em}\mathrm{for}\hspace{0.5em}\alpha = 4,                         \cr
N_\alpha &= \sqrt{N_{\mathrm {\mathrm T}} N_{\mathrm L}}
\hspace{1em}\mathrm{for}\hspace{0.5em}\alpha = (5-8).
\end{align}
We finally obtain the SDMEs in terms of the helicity amplitudes:
\begin{align}
\label{eq:def:SDME2}
\rho_{\lambda\lambda'}^0 &= \frac{1}{2N_{\mathrm T}}
\sum_{\lambda_\gamma = \pm 1}
{\mathcal M}_{\lambda \lambda_\gamma} {\mathcal M}_{\lambda' \lambda_\gamma}^*,       \cr
\rho_{\lambda\lambda'}^1 &= \frac{1}{2N_{\mathrm T}}
\sum_{\lambda_\gamma = \pm 1}
{\mathcal M}_{\lambda -\lambda_\gamma} {\mathcal M}_{\lambda' \lambda_\gamma}^*,      \cr
\rho_{\lambda\lambda'}^2 &= \frac{i}{2N_{\mathrm T}}
\sum_{\lambda_\gamma = \pm 1} \lambda_\gamma
{\mathcal M}_{\lambda -\lambda_\gamma} {\mathcal M}_{\lambda' \lambda_\gamma}^*,      \cr
\rho_{\lambda\lambda'}^3 &= \frac{1}{2N_{\mathrm T}}
\sum_{\lambda_\gamma = \pm 1} \lambda_\gamma
{\mathcal M}_{\lambda \lambda_\gamma} {\mathcal M}_{\lambda' \lambda_\gamma}^*,       \cr
\rho_{\lambda\lambda'}^4 &= \frac{1}{N_{\mathrm L}}
{\mathcal M}_{\lambda 0} {\mathcal M}_{\lambda'0}^*,                            \cr
\rho_{\lambda\lambda'}^5 &= \frac{1}{\sqrt{2N_{\mathrm T}N_{\mathrm L}}}
\sum_{\lambda_\gamma = \pm 1} \frac{\lambda_\gamma}{2}
({\mathcal M}_{\lambda 0} {\mathcal M}_{\lambda' \lambda_\gamma}^* +
{\mathcal M}_{\lambda \lambda_\gamma} {\mathcal M}_{\lambda' 0}^*),                \cr
\rho_{\lambda\lambda'}^6 &= \frac{i}{\sqrt{2N_{\mathrm T}N_{\mathrm L}}}
\sum_{\lambda_\gamma = \pm 1} \frac{1}{2}
({\mathcal M}_{\lambda 0} {\mathcal M}_{\lambda' \lambda_\gamma}^* -
{\mathcal M}_{\lambda \lambda_\gamma} {\mathcal M}_{\lambda' 0}^*),                \cr
\rho_{\lambda\lambda'}^7 &= \frac{1}{\sqrt{2N_{\mathrm T}N_{\mathrm L}}}
\sum_{\lambda_\gamma = \pm 1} \frac{1}{2}
({\mathcal M}_{\lambda 0} {\mathcal M}_{\lambda' \lambda_\gamma}^* +
{\mathcal M}_{\lambda \lambda_\gamma} {\mathcal M}_{\lambda' 0}^*),                \cr
\rho_{\lambda\lambda'}^8 &= \frac{i}{\sqrt{2N_{\mathrm T}N_{\mathrm L}}}
\sum_{\lambda_\gamma = \pm 1} \frac{\lambda_\gamma}{2}
({\mathcal M}_{\lambda 0} {\mathcal M}_{\lambda' \lambda_\gamma}^* -
{\mathcal M}_{\lambda \lambda_\gamma} {\mathcal M}_{\lambda' 0}^*),
\end{align}
where the summation over the incoming and outgoing nucleon helicities are
omitted, for brevity, i.e.,
\begin{align}
\label{eq:for:brevity}
\sum_{\lambda_i, \lambda_f}
{\mathcal M}_{\lambda_f \lambda ; \lambda_i \lambda_\gamma}
{\mathcal M}_{\lambda_f \lambda' ; \lambda_i \lambda_\gamma}^* =
{\mathcal M}_{\lambda \lambda_\gamma} {\mathcal M}_{\lambda' \lambda_\gamma}^*.
\end{align}

\label{eq:TSymmeRela}
We have the following relations:
\begin{align}
\label{eq:RhoSymmeRela}
\rho_{\lambda\lambda'}^\alpha &= (-1)^{\lambda-\lambda'} \rho_{-\lambda-\lambda'}^\alpha,
\hspace{1em} (\alpha = 0,1,4,5,8),                             \cr
\rho_{\lambda\lambda'}^\alpha &= -(-1)^{\lambda-\lambda'} \rho_{-\lambda-\lambda'}^\alpha,
\hspace{1em} (\alpha = 2,3,6,7).
\end{align}



\begin{thebibliography}{99}
\bibitem{Breitweg:1998nh} 
  J.~Breitweg {\it et al.} (ZEUS Collaboration),
  Eur.\ Phys.\ J.\ C {\bf 6}, 603 (1999).

\bibitem{Aaron:2009xp}
  F.~D.~Aaron {\it et al.} (H1 Collaboration),
J.\,High Energy Phys. {\bf 05}, 032 (2010).

\bibitem{Breitweg:2000mu} 
  J.~Breitweg {\it et al.} (ZEUS Collaboration) ,
  Phys.\ Lett.\ B {\bf 487}, 273 (2000).

\bibitem{Adloff:2000nx} 
  C.~Adloff {\it et al.} (H1 Collaboration),
  Phys.\ Lett.\ B {\bf 483}, 360 (2000).
  
\bibitem{Chekanov:2005cqa} 
  S.~Chekanov {\it et al.} (ZEUS Collaboration),
  Nucl.\ Phys.\ {\bf B718}, 3 (2005).

\bibitem{Dixon:1977pg}                                          
  R.~L.~Dixon {\it et al.},
  Phys.\ Rev.\ Lett.\ {\bf 39}, 516 (1977).

\bibitem{Dixon:1978vy}                                          
  R.~L.~Dixon {\it et al.},
  Phys.\ Rev.\ D {\bf 19}, 3185 (1979).

\bibitem{Cassel:1981sx}                                         
  D.~G.~Cassel {\it et al.},
  Phys.\ Rev.\ D {\bf 24}, 2787 (1981).

\bibitem{Hadjidakis:2004zm} 
  C.~Hadjidakis {\it et al.} (CLAS Collaboration),
  Phys.\ Lett.\ B {\bf 605}, 256 (2005).

\bibitem{Morrow:2008ek} 
  S.~A.~Morrow {\it et al.} (CLAS Collaboration),
  Eur.\ Phys.\ J.\ A {\bf 39}, 5 (2009).

\bibitem{Lukashin:2001sh}                                       
  K.~Lukashin {\it et al.} (CLAS Collaboration),
  Phys.\ Rev.\ C {\bf 63}, 065205 (2001); {\bf 64}, 059901 (E) (2001).

\bibitem{Morand:2005ex} 
  L.~Morand {\it et al.} (CLAS Collaboration),
  Eur.\ Phys.\ J.\ A {\bf 24}, 445 (2005).

\bibitem{Santoro:2008ai}                                        
  J.~P.~Santoro {\it et al.} (CLAS Collaboration),
  Phys.\ Rev.\ C {\bf 78}, 025210 (2008).

\bibitem{Borissov:2000wb}                                       
  A.~Borissov (HERMES Collaboration),
  Acta Phys.\ Polon.\ B {\bf 31}, 2353 (2000).

\bibitem{Airapetian:2010dh} 
  A.~Airapetian {\it et al.} (HERMES Collaboration) ,
  Eur.\ Phys.\ J.\ C {\bf 71}, 1609 (2011).

\bibitem{Airapetian:2014gfp} 
  A.~Airapetian {\it et al.} (HERMES Collaboration),
  Eur.\ Phys.\ J.\ C {\bf 74}, 3110 (2014); {\bf 76}, 162 (E) (2016).

\bibitem{Airapetian:2017vit} 
  A.~Airapetian {\it et al.} (HERMES Collaboration),
  Eur.\ Phys.\ J.\ C {\bf 77}, 378 (2017).


\bibitem{Laget:1994ba} 
  J.~M.~Laget and R.~Mendez-Galain,
  Nucl.\ Phys.\ {\bf A581}, 397 (1995).

\bibitem{Laget:2000gj} 
  J.~M.~Laget,
  Phys.\ Lett.\ B {\bf 489}, 313 (2000).

\bibitem{Laget:2001mu} 
  J.~M.~Laget,
  Nucl.\ Phys.\ A {\bf 699}, 184 (2002)

\bibitem{Laget:2004qu} 
  J.~M.~Laget,
  Phys.\ Rev.\ D {\bf 70}, 054023 (2004).

\bibitem{Obukhovsky:2009th}
  I.~T.~Obukhovsky \textit{et al.}, 
  Phys.\ Rev.\ D {\bf 81}, 013007 (2010).
 
\bibitem{Kim:2019kef}
  S.~H.~Kim and S.~i.~Nam,
  Phys.\ Rev.\ C {\bf 100}, 065208 (2019).

\bibitem{Seraydaryan:2013ija}
  H.~Seraydaryan {\it et al.} (CLAS Collaboration),
  Phys.\ Rev.\ C {\bf 89}, 055206 (2014).

\bibitem{Dey:2014tfa} 
  B.~Dey {\it et al.} (CLAS Collaboration),
  Phys.\ Rev.\ C {\bf 89}, 055208 (2014); {\bf 90}, 019901 (E) (2014).

\bibitem{Oh:2003aw} 
  Y.~s.~Oh and T.~S.~H.~Lee,
  Phys.\ Rev.\ C {\bf 69}, 025201 (2004).

\bibitem{Wei:2019imo} 
  N.~C.~Wei, F.~Huang, K.~Nakayama, and D.~M.~Li,
  Phys.\ Rev.\ D {\bf 100}, 114026 (2019).

\bibitem{Donnachie:1987abc}
A.~Donnachie and P.~V.~Landshoff,
  Nucl.\ Phys.\ {\bf B244}, 322 (1984); {\bf B267}, 690 (1986);
  Phys.\ Lett.\ B {\bf 185}, 403 (1987).

\bibitem{Pichowsky:1996jx}
  M.~A.~Pichowsky and T.~S.~H.~Lee,
  Phys.\ Lett.\ B {\bf 379}, 1 (1996).

\bibitem{Pichowsky:1996tn} 
  M.~A.~Pichowsky and T.~S.~H.~Lee,
  Phys.\ Rev.\ D {\bf 56}, 1644 (1997).

\bibitem{Titov:2003bk}
  A.~I.~Titov and T.~S.~H.~Lee,
  Phys.\ Rev.\ C {\bf 67}, 065205 (2003).

\bibitem{Titov:1998bw}
  A.~I.~Titov, Y.~Oh, S.~N.~Yang, and T.~Morii,
  Phys.\ Rev.\ C {\bf 58}, 2429 (1998).

\bibitem{Donnachie:1988nj} 
  A.~Donnachie and P.~V.~Landshoff,
  Nucl.\ Phys.\ {\bf B311}, 509 (1989).

\bibitem{Jaroszkiewicz:1974ep} 
  G.~A.~Jaroszkiewicz and P.~V.~Landshoff,
  Phys.\ Rev.\ D {\bf 10}, 170 (1974).

\bibitem{Donnachie:1983hf}
  A.~Donnachie and P.~V.~Landshoff,
  Nucl.\ Phys.\ {\bf B231}, 189 (1984).

\bibitem{Kochelev:1999zf}
  N.~I.~Kochelev, D.~P.~Min, Y.~Oh, V.~Vento, and A.~V.~Vinnikov,
  Phys.\ Rev.\ D {\bf 61}, 094008 (2000).

\bibitem{Kaiser:1990yf}
  N.~Kaiser and U.~G.~Meissner,
  Nucl.\ Phys.\ {\bf A519}, 671 (1990).

\bibitem{Tanabashi:2018oca} 
  M.~Tanabashi {\it et al.} (Particle Data Group),
  Phys.\ Rev.\ D {\bf 98}, 030001 (2018).

\bibitem{Birkel:1995ct}
  M.~Birkel and H.~Fritzsch,
  Phys.\ Rev.\ D {\bf 53}, 6195 (1996).

\bibitem{Donachie2002}
  A.~Donnachie \textit{et al.}, 
  $Pomeron\, Physics\, and\, QCD$ (Cambridge University Press, Cambridge,
  England, 2002).

\bibitem{Stoks:1999bz}
  V.~G.~J.~Stoks and Th.~A.~Rijken,
  Phys.\ Rev.\ C {\bf 59}, 3009 (1999).

\bibitem{Rijken:1998yy} 
  T.~A.~Rijken, V.~G.~J.~Stoks and Y.~Yamamoto,
  Phys.\ Rev.\ C {\bf 59}, 21 (1999).
  
\bibitem{Titov:1999eu} 
  A.~I.~Titov, T.-S.~H.~Lee, H.~Toki, and O.~Streltsova,
  Phys.\ Rev.\ C {\bf 60}, 035205 (1999).

\bibitem{Meissner:1997qt}
  U.~G.~Meissner, V.~Mull, J.~Speth, and J.~W.~van Orden,
  Phys.\ Lett.\ B {\bf 408}, 381 (1997).

\bibitem{Davidson:2001rk} 
  R.~M.~Davidson and R.~Workman,
  Phys.\ Rev.\ C {\bf 63}, 025210 (2001).

\bibitem{Nam:2017yeg} 
  S.~i.~Nam,
  Phys.\ Rev.\ D {\bf 96}, 076021 (2017).

\bibitem{Kelly:2004hm} 
  J.~J.~Kelly,
  Phys.\ Rev.\ C {\bf 70}, 068202 (2004).
  
\bibitem{Ballam:1972eq}
  J.~Ballam {\it et al.},
  Phys.\ Rev.\ D {\bf 7}, 3150 (1973).
  
\bibitem{Barber:1981fj}
  D.~P.~Barber {\it et al.},
  Z.\ Phys.\ C {\bf 12}, 1 (1982).

\bibitem{Egloff:1979mg}
  R.~M.~Egloff {\it et al.},
  Phys.\ Rev.\ Lett.\ {\bf 43}, 657 (1979).

\bibitem{Schilling:1973ag} 
  K.~Schilling and G.~Wolf,
  Nucl.\ Phys.\ {\bf B61}, 381 (1973).

\bibitem{Rosenbluth:1950yq} 
  M.~N.~Rosenbluth,
  Phys.\ Rev.\ {\bf 79}, 615 (1950).

\bibitem{Tytgat}
  M.~Tytgat, DESY-THESIS-2001-018 (2001).
\end{thebibliography}
\end{document}